\documentclass[aps,pra,reprint,groupedaddress]{revtex4-1}
\usepackage{graphicx}
\usepackage{color}
\graphicspath{{figure/}}
\usepackage{amsmath}
\usepackage{hyperref}
\graphicspath{{figure/}}
\begin{document}

\title{Effect of trap symmetry and atom-atom interactions on a trapped atom interferometer with internal state labelling}

\author{M. Dupont-Nivet$^{1}$\footnote{Corresponding author: matthieu.dupontnivet@thalesgroup.com}, C. I. Westbrook$^{2}$ and S. Schwartz$^{3}$}

\affiliation{${}^{1}$Thales Research and Technology France, 1 av. Fresnel, 91767 Palaiseau, France \\
${}^{2}$Universit\'e Paris-Saclay, Institut d'Optique Graduate School, CNRS, Laboratoire Charles Fabry, 91127 Palaiseau, France  \\
${}^{3}$Laboratoire Kastler Brossel, ENS-Universit\'e PSL, CNRS, Sorbonne Universit\'e, Coll\`ege de France, 24 rue Lhomond, 75005 Paris, France }

\date{\today}

\begin{abstract}
In this paper, we study the dynamics of a trapped atom interferometer with internal state labelling in the presence of interactions.
We consider two situations: an atomic clock in which the internal states remain superposed, and an inertial sensor configuration in which they are separated.
From the  average spin evolution, we deduce the fringe contrast and the phase-shift. 
In the clock configuration, we recover the well-known identical spin rotation effect (ISRE) which can significantly increase the spin coherence time.
We also find that the magnitude of the effect depends on the trap geometry in a way that is consistent with our recent experimental results in a clock configuration \cite{DupontNivet2017b}, where ISRE was not observed. 
In the case of an inertial sensor, we show that despite the spatial separation it is still possible to increase the coherence time by using mean field interactions to counteract asymmetries of the trapping potential.
\end{abstract}

\maketitle

\section{Introduction}

Trapped, cold atom interferometers play an important role in the realization of sensing devices such as 
atomic clocks \cite{Treutlein2004,Deutsch2010,Szmuk2015}, 
accelerometers \cite{Ammar2014,DupontNivet2014,Pelle2013,Alauze2018,Xu2019}, 
gyroscopes \cite{GarridoAlzar2012,Moan2019} 
and magnetometers \cite{Sadgrove2013,Eto2016}. 
In such devices, as compared for example with interferometers using free-falling atoms, the confinement typically results in higher atom densities, hence stronger atom-atom interactions \cite{Grond2010,Abend2016}. 
Interactions typically limit the coherence of the interferometer phase \cite{Schumm2005,Javanainen1997}, but
can also be used to improve performance through squeezing \cite{Jo2007,Berrada2013,Hosten2016,Haas2014b,Barontini2015}. 
In atom chip based atomic clocks \cite{Deutsch2010}, interactions are responsible for the ``identical spin rotation effect" (ISRE) which has led to remarkably long (one minute) spin coherence times  \cite{Kleine2011,Solaro2016,Deutsch2012}.

\begin{figure*}
\centering  \includegraphics[width=0.75\textwidth]{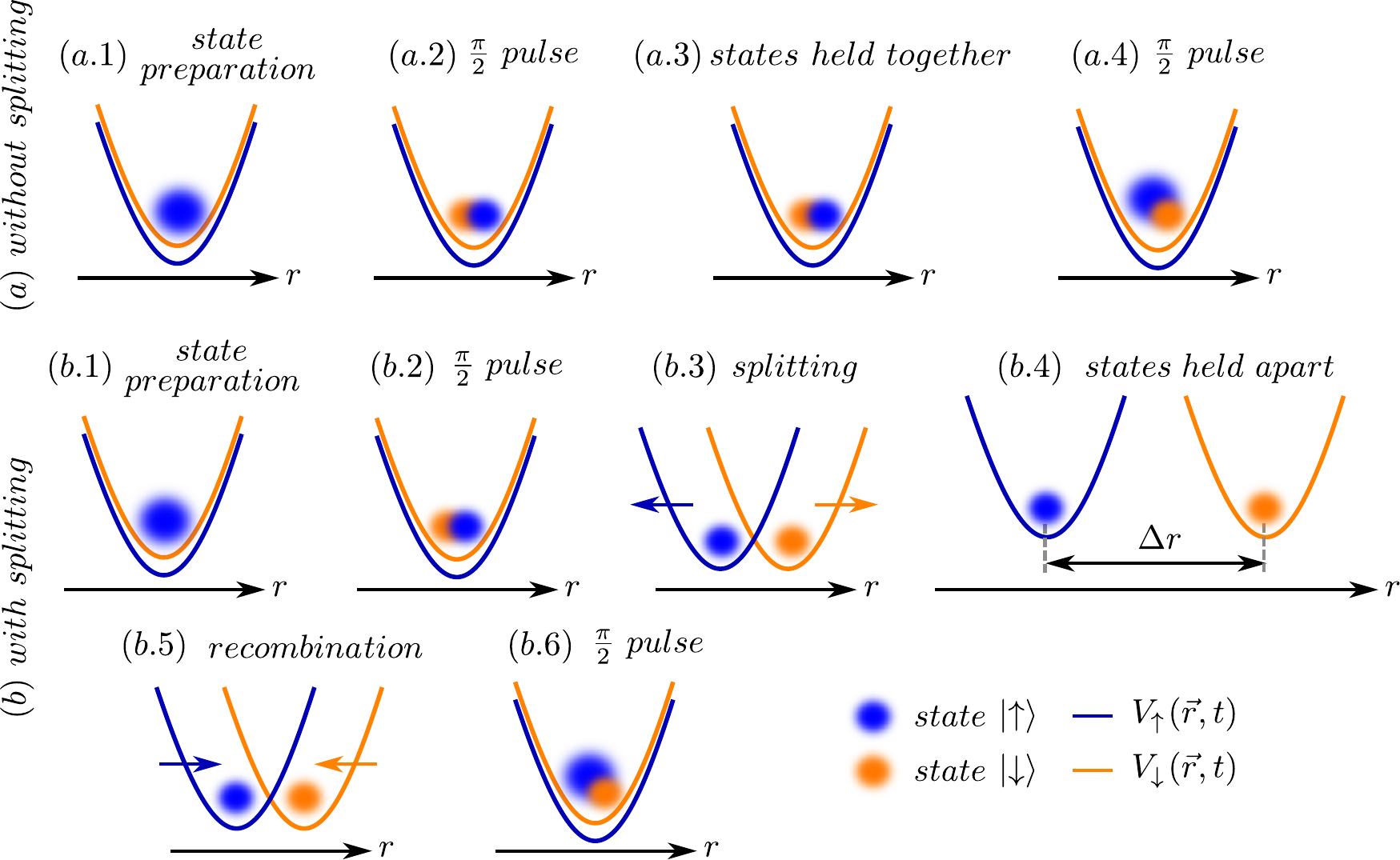}
\caption{\label{fig_inter} (Color online) Schematic diagram of the Ramsey interferometer protocols considered in this article. (a) Atomic clock configuration: internal states are not spatially separated.   (b) Inertial sensor with spatial separation of the two states. In the clock, a typical sequence is as follows: (a.1) the atomic cloud is prepared in internal state $\left|\uparrow\right>$. (a.2) A first $\pi/2$ pulse puts the atoms in a coherent superposition of the two internal states $\left|\uparrow\right>$ and $\left|\downarrow\right>$. (a.3) The two states remain overlapped and are allowed to evolve. (a.4) A second $\pi/2$ pulse closes the interferometer. In the case with spatial separation, a typical sequence is: (b.1) The atomic cloud is prepared in internal state $\left|\uparrow\right>$. (b.2) A first $\pi/2$ pulse puts the atoms in a coherent superposition of the two internal states $\left|\uparrow\right>$ and $\left|\downarrow\right>$. (b.3) The two trapping potentials $V_{\uparrow}(\vec{r},t)$ and $V_{\downarrow}(\vec{r},t)$ spatially separate the two internal states. (b.4) The two internal states evolve while held apart, (b.5) The two clouds are brought together again. (b.6) A second $\pi/2$ pulse closes the interferometer. The size of the blue (orange) shaded-disc, represents the population in state $\left|\uparrow\right>$ ($\left|\downarrow\right>$). The blue (orange) arrows indicate the direction of the displacement of the trap $V_{\uparrow}(\vec{r},t)$ ($V_{\downarrow}(\vec{r},t)$).
}
\end{figure*}

The trapped atom inertial sensors (accelerometers and gyroscopes) described in references \cite{Ammar2014,DupontNivet2014,DupontNivet2016} (see also figure \ref{fig_inter}) resemble atomic clocks in that they depend on the creation of superpositions of different internal states (hereafter noted as $\left|\uparrow\right>$ and $\left|\downarrow\right>$). 
But, unlike clocks, they also require that the two internal states be spatially separated and later recombined \cite{Bohi2009,Ammar2014} (see figure \ref{fig_inter}.b.3). 
Therefore one does not expect the ISRE to be present. 
In such an interferometer, and in the absence of interactions, we have shown that the coherence time, defined by the decay time of the fringe contrast, is governed by the asymmetry in the trapping potentials of the two arms \cite{DupontNivet2014,DupontNivet2017b}. 
One objective of this manuscript is to study how the presence of interactions affects those predictions. 
Another objective is to study the link between ISRE and the geometry of the trapping potential in the clock configuration, motivated by the fact that ISRE was not observed in our recent experiments \cite{DupontNivet2017b} despite the similarity of our apparatus to that of Ref.~\cite{Deutsch2010}. 
We therefore have undertaken a theoretical study of a trapped spinor gas, deriving an equation for the time evolution of the average spin in the presence of atom-atom interactions for several trapping geometries with and without spatial separation of the two internal states.
We have identified the differences between our geometry and that of Ref.~\cite{Deutsch2010} which account for the absence of ISRE in our case~\cite{DupontNivet2017b}.
For the trapped atom inertial sensor, we find that ISRE does not play an important role, as expected. Still, the analysis illustrates a potentially useful effect of interactions when the two spin states are separated: if the traps are not exactly identical, the presence of mean field shifts can be used to partially compensate for the dephasing induced by the trapping potential.

Atom interactions in spin mixtures are known to create spin waves \cite{Laloe1988,Lhuillier1982b,Lhuillier1982a,Bashkin1981,Bashkin1984,Bashkin1986}.
Spin waves have also been observed in cold atom experiments \cite{Lewandowski2002,Du2008,Du2009b}. 
Effects of interactions at low temperature have been extensively studied in gases such as helium and hydrogen. 
For example the work of Bashkin \cite{Bashkin1981,Bashkin1984,Bashkin1986} describes such effects on magnetic and transport properties. 
The work of Lhuillier \cite{Lhuillier1982b,Lhuillier1982a,Bouchaud1985} was one of the first to describe the ISRE, and the observations reported in Refs.~\cite{Deutsch2010,Kleine2011} have stimulated further work \cite{Liu2013b}.
We will follow the approaches developed by these authors in the following.

This paper is organized as follows. 
In section \ref{sec_Introduction}, we describe the Hamiltonian for the one atom average spin that we use to model the system.
The results for the one atom average spin evolution equation are summarized in section \ref{sec_Equation} 
for the clock and for the inertial sensor configurations. 
The derivations of these results are given in appendix \ref{sec_Demonstration}. 
These equations require the computation of the atom-atom interaction kernel, which we do in section \ref{sec_Kernel} for three different interaction geometries: plane waves, a one dimensional harmonic trap and a three dimensional isotropic harmonic trap. 
Section \ref{sec_Numeric} links the one atom average spin to the contrast and the phase-shift of an interferometer. 
We also perform numerical studies of the contrast and phase-shift.
We show that the ISRE in the clock configuration is much less important in a spherical geometry. 
In the case of the inertial sensor, we show how one can actually increase the contrast decay time by using a spin mixing pulse area which is different from  $\pi/2$ pulse, building on mean field interactions in the trap.  
Calculation details are given in appendices \ref{sec_Demonstration}, \ref{sec_Appendix} and \ref{sec_FullISRE}.

\section{Problem definition}
\label{sec_Introduction}

We start by considering an atom in a gas of $N$ identical atoms. 
We label this trial atom by $1$. 
The dynamics of the mean of the spin operator of the trial atom $\vec{S}_1$ is given by the Ehrenfest theorem:
\begin{eqnarray}
\frac{d}{dt} \left< \vec{S}_1 \right> = \frac{i}{\hbar} \left< \left[ \widehat{H},\vec{S}_1 \right] \right> + \left< \frac{\partial \vec{S}_1}{\partial t} \right> ,
\label{eq_SpinOp}
\end{eqnarray}
where $\widehat{H}$ is the Hamiltonian describing our system. 
The one atom spin $\vec{S}_1$ can be written in the basis $\left\{ \left|\uparrow\right>,\left|\downarrow\right> \right\}$ in terms of Pauli matrices:
\begin{eqnarray}
\vec{S}_1 = \frac{1}{2}\left( \sigma_x^1 \vec{e}_x + \sigma_y^1 \vec{e}_y + \sigma_z^1 \vec{e}_z \right) ,
\end{eqnarray}
with:
\begin{eqnarray}
\sigma_x  & = &
\begin{pmatrix}
0&  1\\
1& 0
\end{pmatrix} ,
\qquad
\sigma_y =
\begin{pmatrix}
0&  -i\\
i&  0 
\end{pmatrix} , \nonumber\\
\sigma_z & = &
\begin{pmatrix}
1&  0\\
0&  -1
\end{pmatrix} ,
\qquad
Id =
\begin{pmatrix}
1&  0\\
0&  1
\end{pmatrix} .
\end{eqnarray}

\subsection{Model for the Hamiltonian}
\label{sec_HModele}

We write the Hamiltonian of our system of $N$ atoms trapped in the state dependent potentials $V_\uparrow(\vec{r})$ and $V_\downarrow(\vec{r})$ as:
\begin{eqnarray}
\widehat{H} = \widehat{H}_0^{mean} + \widehat{H}_0^{diff} + \widehat{H}_{int} ,
\label{eq_H_gene}
\end{eqnarray}
where:
\begin{eqnarray*}
\widehat{H}_0^{mean} + \widehat{H}_0^{diff} & = & \frac{\widehat{p}^2}{2m} \left[ \left|\uparrow\right>\left<\uparrow\right|+\left|\downarrow\right>\left<\downarrow\right| \right] \nonumber\\
& + & V_\uparrow(\widehat{\vec{r}})\left|\uparrow\right>\left<\uparrow\right| + V_\downarrow(\widehat{\vec{r}})\left|\downarrow\right>\left<\downarrow\right|
\end{eqnarray*}
corresponds to the total energy of one atom without taking into account the interactions. 
The operators $\widehat{\vec{p}}$ and $\widehat{\vec{r}}$ correspond the momentum and position respectively.
We suppose that the two trapping potentials $V_\uparrow(\vec{r})$ and $V_\downarrow(\vec{r})$ are harmonic but slightly different, so that the vibrational frequencies are not equal:
 $\omega_{\uparrow, \downarrow}^j = \omega_j \pm \delta \omega_j /2$, $j$ stands for the space coordinate ($j=\{x,y,z\}$).
We define $\mathcal{E}_n=\sum_j\hbar \omega_j (n_j+1/2)$ and $\Omega(\mathcal{E}_n)=\sum_j\delta \omega_j (n_j+1/2)$, where $n=(n_x,n_y,n_z)$.
To simplify the discussion we will suppose that the two potentials are similar enough that the vibrational eigenstates, noted $\left|\phi_{\mathcal{E}_n}\right>$, can be considered to be the same in the two wells. 
This assumption is lifted in appendix  \ref{sec_FullISRE}.
We can then write the first two terms of the Hamiltonian as:
\begin{eqnarray}
\widehat{H}_0^{mean} & = & \sum_\mathcal{E} \left( \left|\phi_\mathcal{E}\right>\left|\uparrow\right>\left<\uparrow\right|\left<\phi_\mathcal{E}\right| +  \left|\phi_\mathcal{E}\right>\left|\downarrow\right>\left<\downarrow\right|\left<\phi_\mathcal{E}\right|  \right) \nonumber\\
& = & \sum_\mathcal{E} \mathcal{E} \left|\phi_\mathcal{E}\right> Id \left<\phi_\mathcal{E}\right| ,
\end{eqnarray}
where we discard the subscript $n$ to simplify the notation, and:

\begin{eqnarray}
\widehat{H}_0^{diff} & = & \sum_\mathcal{E} \frac{\hbar\Omega(\mathcal{E})}{2} \left( \left|\phi_\mathcal{E}\right>\left|\uparrow\right>\left<\uparrow\right|\left<\phi_\mathcal{E}\right| - \left|\phi_\mathcal{E}\right>\left|\downarrow\right>\left<\downarrow\right| \left<\phi_\mathcal{E}\right| \right) \nonumber\\
& = & \sum_\mathcal{E} \frac{\hbar\Omega(\mathcal{E})}{2} \left|\phi_\mathcal{E}\right> \sigma_z \left<\phi_\mathcal{E}\right| .
\end{eqnarray}
The sum $\sum_\mathcal{E}$ runs over all states $\left|\phi_\mathcal{E}\right>$ and can include degeneracies.

Under the assumption that $\delta\omega/\omega\simeq\delta\omega_j/\omega_i$ for $j=x,y,z$, the energy difference between the states $\left|\phi_\mathcal{E}\right>\left|\uparrow\right>$ and $\left|\phi_\mathcal{E}\right>\left|\downarrow\right>$ can be written as:
\begin{equation}
\Omega(\mathcal{E})=\frac{\mathcal{E}}{k_BTt_c}
\end{equation}
where we have assumed the atoms are at a temperature $T$ and introduced a coherence time \cite{DupontNivet2014}:
\begin{equation}
t_c=\frac{1}{\delta\omega}\frac{\hbar\omega}{k_BT} .
\end{equation}
Here $k_B$ is the Boltzmann constant, 
and we will use $E$ to denote the energy in units of $k_BT$: $E=\mathcal{E}/k_BT$.

The third term of the Hamiltonian (\ref{eq_H_gene}) corresponds to interactions of the trial atom with the other atoms of the gas. 
At the temperatures we are considering, these interactions can be described entirely by s-wave collisions whose scattering lengths will be denoted as $a_{\uparrow\uparrow}$,  $a_{\uparrow\downarrow}$ and $a_{\downarrow\downarrow}$. 
The interaction Hamiltonian thus reduces to:
\begin{widetext}
\begin{eqnarray}
\widehat{H}_{int} & = & \frac{4\pi\hbar^2a_{\uparrow\downarrow}}{m} \frac{1}{2} \sum_{E_1,E_2,E_3} I^{E_1+E_3,E_2-E_3}_{E_1,E_2} \sum_{j=2}^N \times \left\{ \left|\phi_{E_1+E_3}\right>_1\left|\uparrow\right>_1 {}_{1} \! \left<\downarrow\right| {}_{1} \! \left<\phi_{E_1}\right| \otimes \left|\phi_{E_2-E_3}\right>_j\left|\downarrow\right>_j {}_{j} \! \left<\uparrow\right| {}_{j} \! \left<\phi_{E_2}\right| \right.  \nonumber\\
& & \hspace{5.46cm} \quad + \left. \left|\phi_{E_1+E_3}\right>_1\left|\downarrow\right>_1 {}_{1} \! \left<\uparrow\right| {}_{1}\! \left<\phi_{E_1}\right| \otimes \left|\phi_{E_2-E_3}\right>_j\left|\uparrow\right>_j {}_{j} \! \left<\downarrow\right| {}_{j}\! \left<\phi_{E_2}\right| \right.  \nonumber\\
& & \hspace{5.46cm} \quad + \left. \left|\phi_{E_1+E_3}\right>_1\left|\uparrow\right>_1 {}_{1} \! \left<\uparrow\right| {}_{1}\! \left<\phi_{E_1}\right| \otimes \left|\phi_{E_2-E_3}\right>_j\left|\downarrow\right>_j {}_{j} \! \left<\downarrow\right| {}_{j}\! \left<\phi_{E_2}\right| \right.  \nonumber\\
& & \hspace{5.46cm} \quad + \left. \left|\phi_{E_1+E_3}\right>_1\left|\downarrow\right>_1 {}_{1} \! \left<\downarrow\right| {}_{1} \! \left<\phi_{E_1}\right| \otimes \left|\phi_{E_2-E_3}\right>_j\left|\uparrow\right>_j {}_{j} \! \left<\uparrow\right| {}_{j} \! \left<\phi_{E_2}\right| \right\}  \nonumber\\
& + & \frac{4\pi\hbar^2a_{\uparrow\uparrow}}{m} \sum_{E_1,E_2,E_3} I^{E_1+E_3,E_2-E_3}_{E_1,E_2} \sum_{j=2}^N  \left\{ \left|\phi_{E_1+E_3}\right>_1\left|\uparrow\right>_1 {}_{1} \! \left<\uparrow\right| {}_{1} \! \left<\phi_{E_1}\right| \otimes \left|\phi_{E_2-E_3}\right>_j\left|\uparrow\right>_j {}_{j} \! \left<\uparrow\right| {}_{j} \! \left<\phi_{E_2}\right| \right\}  \nonumber\\
& + & \frac{4\pi\hbar^2a_{\downarrow\downarrow}}{m} \sum_{E_1,E_2,E_3} I^{E_1+E_3,E_2-E_3}_{E_1,E_2} \sum_{j=2}^N\left\{ \left|\phi_{E_1+E_3}\right>_1\left|\downarrow\right>_1 {}_{1} \! \left<\downarrow\right| {}_{1} \! \left<\phi_{E_1}\right| \otimes \left|\phi_{E_2-E_3}\right>_j\left|\downarrow\right>_j {}_{j} \! \left<\downarrow\right| {}_{j} \! \left<\phi_{E_2}\right| \right\} .
\label{eq_HInt}
\end{eqnarray}
$I^{E_1+E_3,E_2-E_3}_{E_1,E_2}$ is the overlap of the wave functions:
\begin{eqnarray}
I^{E_1+E_3,E_2-E_3}_{E_1,E_2} = \int \phi_{E_1+E_3}^*(r)\phi_{E_2-E_3}^*(r)\phi_{E_2}(r)\phi_{E_1}(r) dr .
\label{eq_WaveOverLap}
\end{eqnarray}
\end{widetext}

As stated above, we assume 
$\phi_E^\uparrow(r)=\phi_E^\downarrow(r)$, thus in equation (\ref{eq_WaveOverLap}), we dropped the spin index of the atoms involved in the collisions. In appendix \ref{sec_FullISRE}, we give a more general result in the case $\phi_E^\uparrow(r) \neq \phi_E^\downarrow(r)$. As we will see in section \ref{sec_Kernel}, $I^{E_1+E_3,E_2-E_3}_{E_1,E_2}$ contains information about the interaction geometry.


\subsection{Form of the density operator}
\label{sec_Model_density_op}

We consider a thermal gas described by a Boltzmann distribution $e^{-E}$.
If this gas is trapped in an isotropic, three dimensional harmonic trap and if $k_BT/\hbar\omega \gg 1$, then the density of states is given approximately by  $(k_BT/\hbar\omega)^3(E^2/2)$.
The case of an anisotropic harmonic trap is discussed in reference \cite{Kirsten1996}, and other densities of states can be used by replacing $E^2/2$ with the appropriate terms. 
We write the one atom density operator as:
\begin{eqnarray}
\hspace{-2cm} \widehat{\rho} & = & \sum_{E} e^{-E} \left|\phi_E\right>\left|\uparrow\right>\left<\uparrow\right|\left<\phi_E\right| \nonumber\\
& = & \int dE \frac{E^2}{2} e^{-E} \left|\phi_E\right>\left|\uparrow\right>\left<\uparrow\right|\left<\phi_E\right| .
\label{eq_DensityOp}
\end{eqnarray}
The one atom density operator is normalized to have unit trace. The effect of a $\pi/2$ pulse is modelled by: $\left|\uparrow\right>\rightarrow(\left|\uparrow\right>-i\left|\downarrow\right>)/\sqrt(2)$ and $\left|\downarrow\right>\rightarrow(\left|\downarrow\right>-i\left|\uparrow\right>)/\sqrt(2)$. Unlike in reference \cite{DupontNivet2014}, the effect of the phase of the $\pi/2$ pulse is not taken into account because its does not dependent on the atom density and we are only interested in the effect of the atom-atom interaction on the contrast decay and the phase-shift. After the first $\pi/2$ pulse, the density operator is:
\begin{eqnarray}
\widehat{\rho} = \frac{1}{2} \sum_{E} e^{-E} \left|\phi_E\right>\left( Id - \sigma_y \right)\left<\phi_E\right| .
\label{eq_DensityOp_t0}
\end{eqnarray}

\subsection{Definition of the mean}

We define the one atom average spin at the energy $E$ of the trial atom by writing the trace reduced to the subspace $\left|\phi_E\right>\left<\phi_E\right|$. For an operator $\widehat{X}$ it can be written as:
\begin{eqnarray}
\left<\widehat{X}(E)\right> = \sum_{E'} \left<\phi_{E'}\right| \left( \left|\phi_E\right>\left<\phi_E\right| \left< \widehat{\rho} \widehat{X} \right>_s \right) \left|\phi_{E'}\right> .
\label{eq_OneAtomAverage}
\end{eqnarray}
where $\left<\cdot\right>_s$ is the mean over the spin space. The equation (\ref{eq_SpinOp}) can be rewritten as:
\begin{eqnarray}
\frac{d}{dt} \left<\vec{S}_1(E)\right> = \frac{i}{\hbar} \left< \left[\widehat{H},\vec{S}_1(E)\right]\right> + \left< \frac{\partial}{\partial t} \vec{S}_1(E) \right> .
\label{eq_Spin_Average}
\end{eqnarray} 
Here, $\vec{S}_1$ is a function only of the energy $E$ because we limit our investigation to a regime where an atom oscillates many times in the trap before a collision (the ``collisionless" regime). 


\section{Complete equation for the one atom average spin}
\label{sec_Equation}

\subsection{Case without spatial separation}

Using equations (\ref{eq_AveSpinWithoutDamping}) and (\ref{eq_SpinDamping}) from the appendix,
and defining $\left<\vec{S}_{1}(E)\right>=e^{-E}\left<\vec{\chi}_{1}(E)\right>$,
we obtain the complete equation of motion for the one atom average spin in the absence of spatial separation of internal states:
\begin{eqnarray}
& &\frac{d}{dt} \left<\vec{\chi}_1(E)\right> = \left| \begin{matrix}
0 \\
0 \\
\Omega(E) \end{matrix} \right. \wedge \left<\vec{\chi}_{1}(E)\right> \nonumber\\
& + & \frac{4\pi\hbar aN}{m} \int dE' \frac{E'^2}{2}e^{-E'} I^{E,E'}_{E,E'} \left<\vec{\chi}_{1}(E')\right> \wedge \left<\vec{\chi}_{1}(E)\right>  \nonumber\\
& - & \frac{1}{\tau_{th}} \left( \left<\vec{\chi}_{1}(E)\right> - \int dE \frac{E^2}{2}e^{-E} \left<\vec{\chi}_{1}(E)\right>  \right) ,
\label{eq_FullSpinDynamics}
\end{eqnarray}
with the thermal relaxation time:
\begin{eqnarray}
\frac{1}{\tau_{th}} = \frac{32}{3}\sqrt{\pi}a^2\overline{n}v_r .
\label{eq_tau}
\end{eqnarray}
We have expressed the sums over the energy as integrals over a density of states in an isotropic three dimensional harmonic trap.
The first line of this equation describes the well-known result: between the two $\pi/2$ pulses of a Ramsey interferometer, the spin rotates in the equatorial plane  around the vertical axis of the Bloch sphere at a rate proportional to the energy difference between the states $\left|\phi_E\right>\left|\uparrow\right>$ and $\left|\phi_E\right>\left|\downarrow\right>$. Here, the frequency of the $\pi/2$ pulse and its detuning from the $\left|\uparrow\right> \leftrightarrow \left|\downarrow\right>$ transition are not considered.
The second line describes the interaction-induced rotation of the one atom average spin at energy $E$ around its mean value over the energy weighted by the wave function overlap. This is the identical spin rotation effect \cite{Deutsch2010,Lhuillier1982a,Lhuillier1982b}.
The last line describes the collisional relaxation of the spin.
This equation has been used for example in \cite{Deutsch2010} to fit the contrast decay of a trapped rubidium clock in presence of identical spin rotation effect.

The second line of equation (\ref{eq_FullSpinDynamics}) for the evolution of the one atom average spin takes the form of a pure rotation if the three scattering lengths are equal and $\phi_E^{\uparrow}(r)=\phi_E^{\downarrow}(r)$. 

\subsection{Case with spatial separation}

If the two spin states are spatially separated during the hold time, the identical spin rotation effect is absent and the spin equation takes a simpler form. 
Equations from appendix \ref{sec_AppendixSplit} lead to the following expression for the one atom average spin:
\begin{eqnarray}
& & \frac{d}{dt} \left<\vec{\chi}_1(E)\right> = \left| \begin{matrix}
0 \\
0 \\
\Omega(E) \end{matrix} \right. \wedge \left<\vec{\chi}_{1}(E)\right>  \nonumber\\
& + & \frac{8\pi\hbar aN}{m} \int dE' \frac{E'^2}{2}e^{-E'} I^{E,E'}_{E,E'} \left| \begin{matrix}
0 \\
0 \\
\left<\chi_{1z}(E')\right> \end{matrix} \right. \wedge \left<\vec{\chi}_{1}(E)\right>  \nonumber\\
& - & \frac{1}{\tau_{th}} \left( \left<\vec{\chi}_{1}(E)\right> - \int dE \frac{E^2}{2}e^{-E} \left<\vec{\chi}_{1}(E)\right>  \right) .
\label{eq_SplittingSpinDynamics}
\end{eqnarray}
We assume that the time necessary to separate the spins is short compared to the contrast decay time (see reference \cite{DupontNivet2014}), therefore we neglect the identical spin rotation effect while the spins are in contact during the separation (figure \ref{fig_inter}.b.3) and the recombination (figure \ref{fig_inter}.b5) stages of the interferometer. 

The results of this paper are not limited to the case of rubidium 87, appendix \ref{sec_Appendix} extends the results to other atomic species in which $a_{\uparrow\downarrow} \neq a_{\uparrow\uparrow} \neq a_{\downarrow\downarrow}$, and appendix \ref{sec_FullISRE} considers the case when the three interaction lengths are different and $\phi_E^{\uparrow}(r)\neq\phi_E^{\downarrow}(r)$. 


\section{Interaction kernel $K(E,E')$}
\label{sec_Kernel}

To perform numerical studies of the evolution of the one atom average spin and show how its evolution varies with the trapping geometry, we need $I_{E,E'}^{E,E'}$ as an explicit function of energy. 
To this purpose we define the interaction kernel as:
\begin{eqnarray}
I_{E,E'}^{E,E'} = \frac{K(E,E')}{V_{eff}}
\end{eqnarray}
where $V_{eff}$ is a volume and $K(E,E')$ is dimensionless.

Three different interaction geometries will be considered: i)~a free gas, ii)~a gas trapped in a one dimensional harmonic potential and iii)~a gas trapped in a three dimensional harmonic isotropic potential. 

\subsection{Free gas}

The simplest example is a free gas in a box with an effective volume $V_{eff}$. 
The atom wavefunctions are plane-waves:
\begin{eqnarray}
\phi_{E}(r) = \frac{1}{\sqrt{V_{eff}}} e^{i\vec{k}\vec{r}} \qquad \text{with:} \qquad k \propto \sqrt{E} ,
\end{eqnarray}
leading to:
\begin{eqnarray}
I_{E,E'}^{E,E'} = \frac{1}{V_{eff}} ,
\end{eqnarray}
The interaction kernel is $K(E,E')=1$, and was used for example in reference \cite{Deutsch2010}.

\subsection{One dimensional trap}

References \cite{Deutsch2010,Kleine2011,Du2009b} discussed the case of a gas trapped in a cigar shaped harmonic potential with transverse and confinement frequencies of $\omega_\perp$ and $\omega_\parallel$, and with $\omega_\perp\gg\omega_\parallel$, under the assumption that only the ground state is populated. In the axial direction we use the WKB approximation \cite{Miller1953,Sakurai1995,Schiff1968} for the wave function, leading to $\phi_E(x,y,z)=g(x)g(y)f_E(z)$ with:
\begin{eqnarray}
& & g(x) = \left( \frac{m\omega_\perp^2}{\pi k_B T} \right)^{1/4} \exp\left( -\frac{m\omega_\perp^2}{2k_B T}x^2 \right) , \\
& & f_E(z) = \left(\frac{m}{2}\right)^{1/4}\sqrt{\frac{\omega}{\pi}} \frac{1}{\left(k_BTE - \frac{m\omega_\parallel^2}{2}z^2\right)^{1/4}}  \nonumber\\
& & \qquad \times \exp\left\{ \pm \frac{i}{\hbar}\int dz \sqrt{2m\left(k_BTE - \frac{m\omega_\parallel^2}{2}z^2\right)} \right. \nonumber\\
& & \qquad \qquad \left. -\frac{i}{\hbar}k_BTEt \right\} ,
\end{eqnarray}
where $E$ is still in units of $k_BT$. 
After some straightforward integration one finds:
\begin{eqnarray}
I_{E,E'}^{E,E'} & = & \frac{K(E,E')}{V_{eff}} , \nonumber
\end{eqnarray}
with:
\begin{eqnarray}
\frac{1}{V_{eff}} = \left( \frac{m}{2\pi k_BT} \right)^{3/2} \omega_\perp\omega_\perp\omega_\parallel ,   \nonumber
\end{eqnarray}
and
\begin{eqnarray}
& & K(E,E') = \nonumber\\
& & \frac{1}{\pi^{3/2}E_m^{1/2}} \int_{-\pi/2}^{\pi/2} \frac{d\theta}{\left(  \frac{\left|E-E'\right|}{E_m} + \cos^2\theta \right)^{1/2}} ,
\end{eqnarray}
where $E_m = \min(E,E')$ and $V_{eff}$ is the effective volume of the harmonic trap \cite{Walraven2010,Du2009b}.
We find the interaction kernel $K(E,E')$ used in reference \cite{Du2009b}. 
In the last equation, $K(E,E)$ is undefined for $E=E'$. 
However $I_{E,E}^{E,E}$ is well defined and finite. Because $\left<\vec{\chi}_{1}(E)\right> \wedge \left<\vec{\chi}_{1}(E)\right>=0$, we will add the condition $K(E,E)=0$ for the numerical studies of section \ref{sec_Numeric}. 
 
\subsection{Three dimensional trap}
\label{subsec_Kernel3D}

In the three dimensional case the WKB approximation can be extended \cite{VanVleck1928,Schiller1962,Schiller1962b,VanHorn1967,Sergeenko2000}, but is cumbersome to use. 
We rather approximate the wave function $\phi_E(r)$ with the product of one dimensional WKB wave functions along each Cartesian axis:
\begin{eqnarray}
\phi_E(x,y,z) & = &  f_{E/3}(x) f_{E/3}(y) f_{E/3}(z).
\end{eqnarray}
We have assumed an isotropic partition of the energy between the Cartesian axes. 
After some straightforward integration, we obtain:
\begin{eqnarray}
I_{E,E'}^{E,E'} = \frac{K(E,E')}{V_{eff}} ,  \nonumber
\end{eqnarray}
with:
\begin{eqnarray}
\frac{1}{V_{eff}} = \left( \frac{m}{2\pi k_BT} \right)^{3/2} \omega^3 ,  \nonumber
\end{eqnarray}
and
\begin{eqnarray}
& & K(E,E') = \nonumber \\
& & \frac{3^{3/2}}{\pi^{9/2}E_m^{3/2}} \left[ \int_{-\pi/2}^{\pi/2} \frac{d\theta}{\left(  \frac{\left|E-E'\right|}{E_m} + \cos^2\theta \right)^{1/2}} \right]^{3} .
\label{eq_3DKernel}
\end{eqnarray}


\section{Numerical studies}
\label{sec_Numeric}

\begin{figure*}
\centering  \includegraphics[width=1\textwidth]{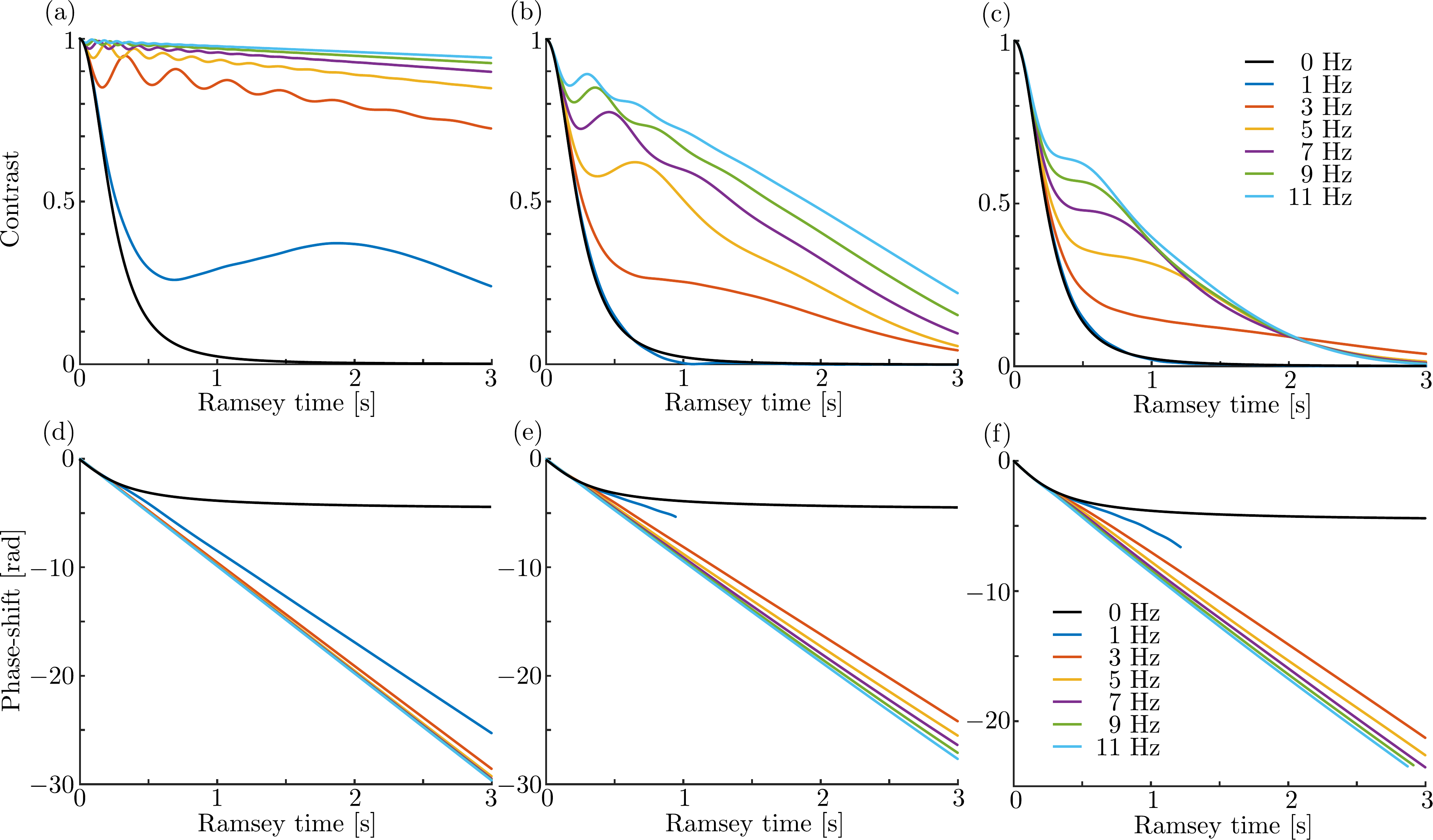}
\caption{\label{fig_ISRE_case_1} (Color online). Case without spatial separation of the two internal states. Contrast (a,b,c) and unwrapped phase-shift [rad] (d,e,f) as a function of the Ramsey time [s], for different atomic densities. The numerical parameters are $\Omega(E)=E/t_c $, $t_c = 300~\text{ms}$, $2\hbar a N /(mV_{eff})=$~$[0\text{ }1\text{ }3\text{ }5\text{ }7\text{ }9\text{ }11]$~Hz and $1/\tau_{th}=0.27\times[0\text{ }1\text{ }3\text{ }5\text{ }7\text{ }9\text{ }11]$~s$^{-1}$. (a,d) plane wave case, (b,e) one dimensional harmonic trap case and (c,f) three dimensional isotropic harmonic trap case. The solid black curve is the case without interaction, namely $2\hbar a N /(mV_{eff})=$~0~Hz and $1/\tau_{th}=$~0~s$^{-1}$ (equation (\ref{eq_StotWithoutInt})). }
\end{figure*}

\begin{figure}
\centering  \includegraphics[width=0.35\textwidth]{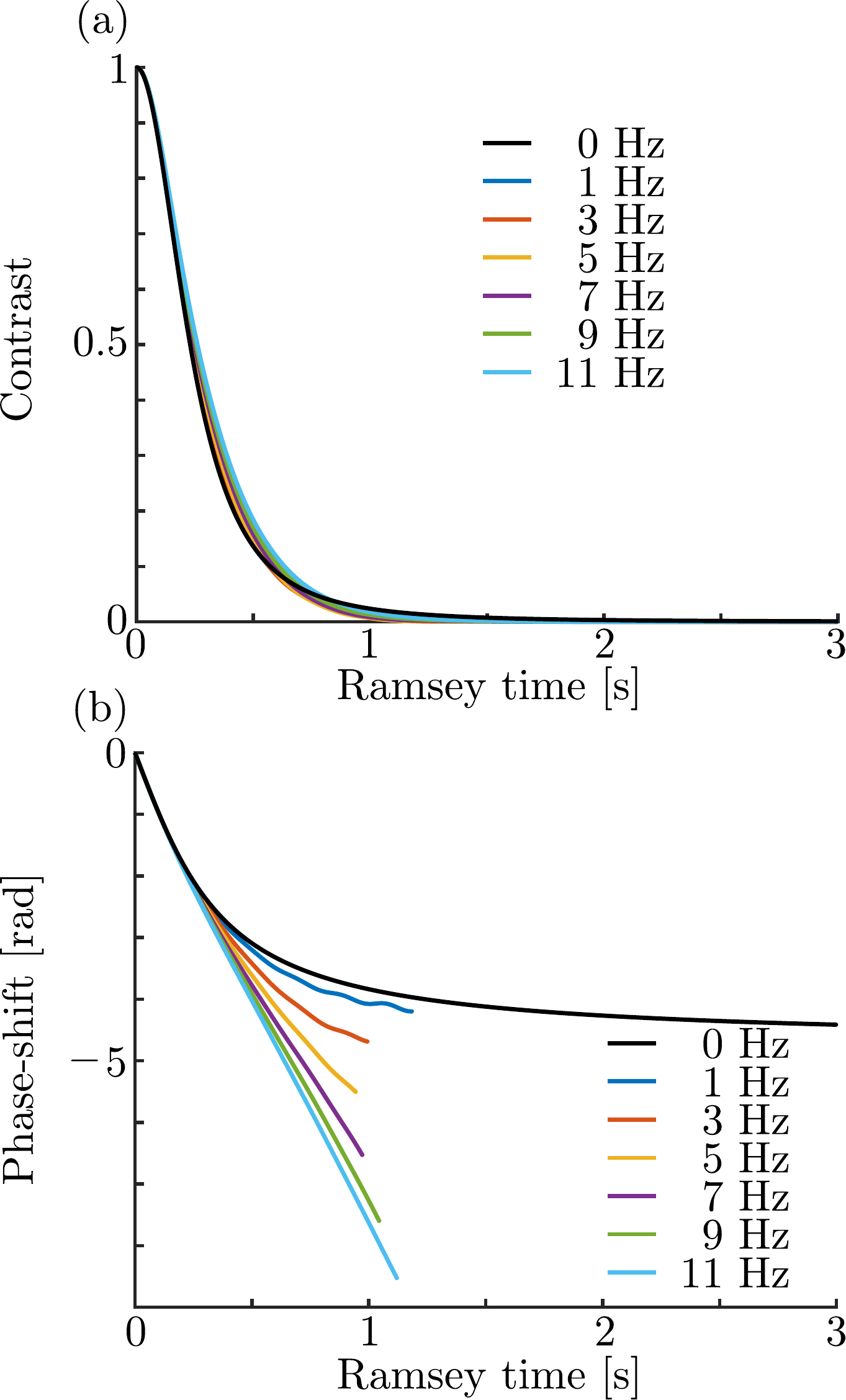}
\caption{\label{fig_ISRE_case_2} (Color online). Case with spatial separation of the two internal states. Contrast (a) and unwrapped phase [rad] (b) as a function of the Ramsey time [s], for different atomic densities. The numerical parameters are $\Omega(E)=E/t_c $, $t_c = 300~\text{ms}$, $2\hbar a N /(mV_{eff})=$~$[0\text{ }1\text{ }3\text{ }5\text{ }7\text{ }9\text{ }11]$~Hz and $1/\tau_{th}=0.27\times[0\text{ }1\text{ }3\text{ }5\text{ }7\text{ }9\text{ }11]$~s$^{-1}$. The plane wave case, one dimensional harmonic trap case and three dimensional isotropic harmonic trap case give similar curves. The solid black curve is the case without interaction, namely $2\hbar a N /(mV_{eff})=$~0~Hz and $1/\tau_{th}=$~0~s$^{-1}$ (equation (\ref{eq_StotWithoutInt})). }
\end{figure}

\begin{figure*}
\centering  \includegraphics[width=0.67\textwidth]{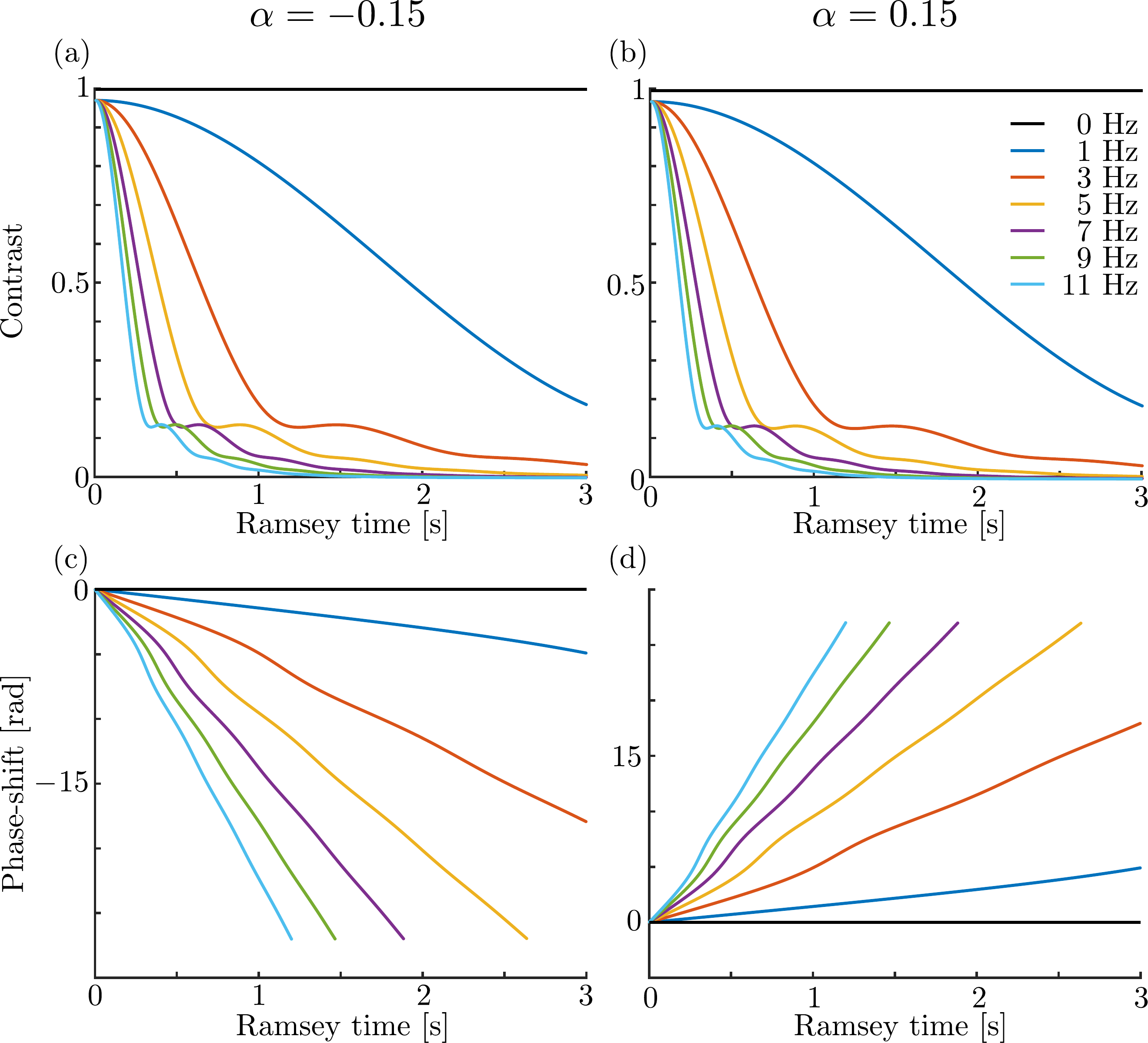}
\caption{\label{fig_MeanField} (Color online). Case with spatial separation of the two internal states, unperfect $\pi/2$ pulses and no asymmetry between the two traps, i.e. $\Omega(E)=0$. Contrast (a,b,) and unwrapped phase-shift [rad] (c,d) for a three dimensional traps (interaction kernel of section \ref{subsec_Kernel3D}) as a function of the Ramsey time [s], for different atomic densities. The numerical parameters are $\Omega(E)=0$, $\delta'(E)=E/(\Omega_R t_c)$, $\Omega_R = 2\pi \times 500~\text{Hz}$, $t_c = 300~\text{ms}$, $2\hbar a N /(mV_{eff})=$~$[0\text{ }1\text{ }3\text{ }5\text{ }7\text{ }9\text{ }11]$~Hz and $1/\tau_{th}=0.27\times[0\text{ }1\text{ }3\text{ }5\text{ }7\text{ }9\text{ }11]$~s$^{-1}$. (a,d) $\alpha=-0.15$, (b,e) $\alpha=0$ and (c,f) $\alpha=0.15$. The solid black curve is the case without interaction, namely $2\hbar a N /(mV_{eff})=$~0~Hz and $1/\tau_{th}=$~0~s$^{-1}$ (equation (\ref{eq_StotWithoutInt})). }
\end{figure*}

\begin{figure*}
\centering  \includegraphics[width=1\textwidth]{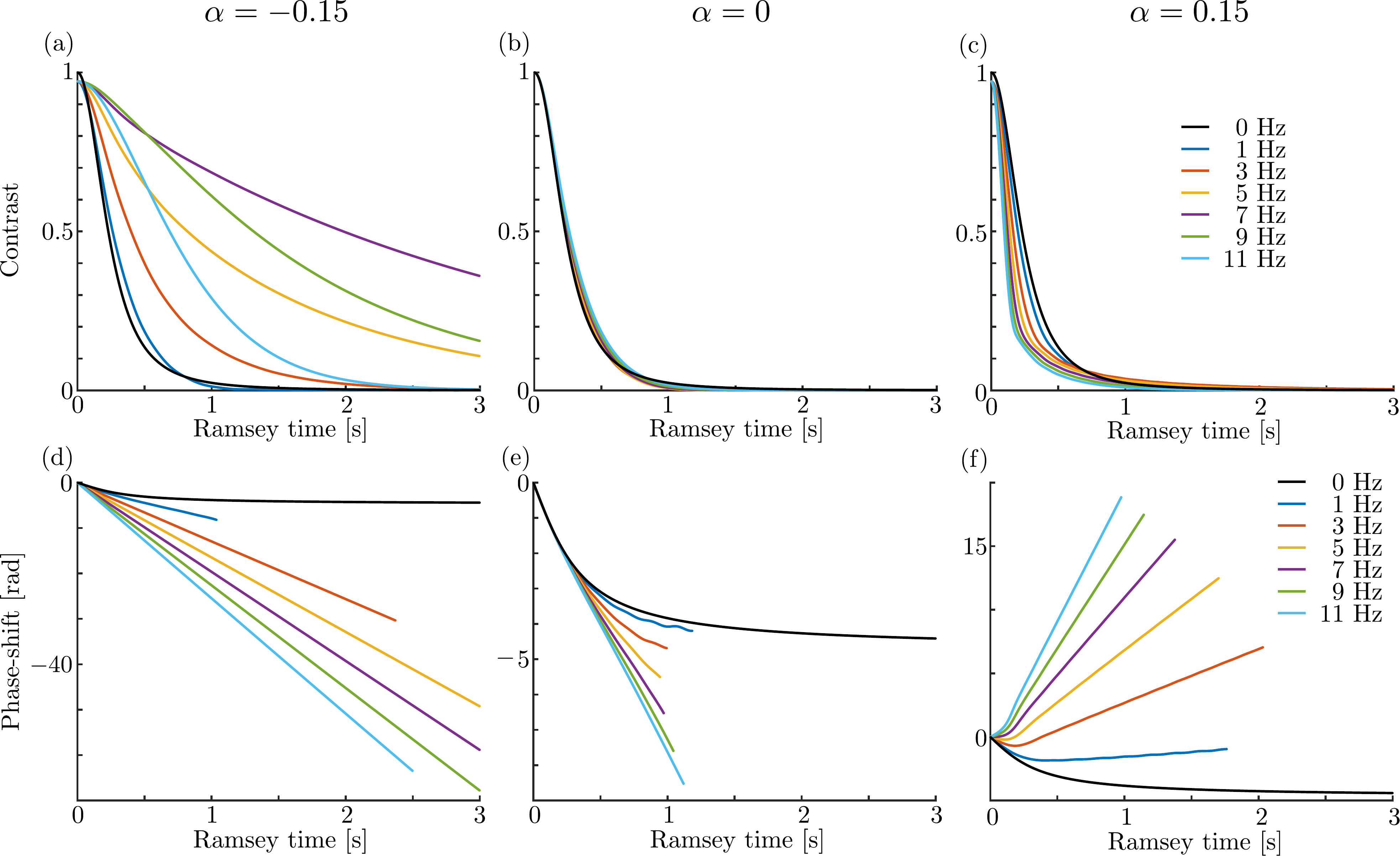}
\caption{\label{fig_ISRE_case_2_enhanced} (Color online). Case with spatial separation of the two internal states and imperfect $\pi/2$ pulses. Contrast (a,b,c) and unwrapped phase-shift [rad] (d,e,f) for a three dimensional traps (interaction kernel of section \ref{subsec_Kernel3D}) as a function of the Ramsey time [s], for different atomic densities. The numerical parameters are $\Omega(E)=E/t_c $, $\delta'(E)=E/(\Omega_R t_c)$, $\Omega_R = 2\pi \times 500~\text{Hz}$, $t_c = 300~\text{ms}$, $2\hbar a N /(mV_{eff})=$~$[0\text{ }1\text{ }3\text{ }5\text{ }7\text{ }9\text{ }11]$~Hz and $1/\tau_{th}=0.27\times[0\text{ }1\text{ }3\text{ }5\text{ }7\text{ }9\text{ }11]$~s$^{-1}$. (a,d) $\alpha=-0.15$, (b,e) $\alpha=0$ and (c,f) $\alpha=0.15$. The solid black curve is the case without interaction, namely $2\hbar a N /(mV_{eff})=$~0~Hz and $1/\tau_{th}=$~0~s$^{-1}$ (equation (\ref{eq_StotWithoutInt})). }
\end{figure*}

We now apply the previous results to the interferometer described in the introduction and in figure \ref{fig_inter}.
We will derive two characteristics of the interferometer: the contrast and the phase-shift. 
The contrast decay time determines how long the phase can be accumulated and thus what sensitivity can be ultimately reached. 
The phase-shift result from the cumulated effect of interactions and potential difference precession rate $\Omega(E)$. 
This quantity can be compared to the ``useful" part of the phase-shift (the one containing the quantity being measured) to determine at what level parameters such as the density and potential difference precession rate need to be stabilized.

Equation (\ref{eq_DensityOp_t0}) gives the density operator just after the first $\pi/2$ pulse. 
To find the contrast and the phase-shift in terms of the components of the one-atom average spin we need an expression for the density operator at a time $t$ after the first $\pi/2$ pulse. This expression is given by \cite{Gamble2009}:
\begin{eqnarray}
\widehat{\rho}(t) = \int dE \frac{E^2}{2}e^{-E} \left|\phi_E\right> \widehat{M} \left<\phi_E\right| ,
\end{eqnarray}
with:
\begin{eqnarray}
\widehat{M} = \frac{Id + \left<\chi_{1x}(E)\right>\sigma_x + \left<\chi_{1y}(E)\right>\sigma_y + \left<\chi_{1z}(E)\right>\sigma_z}{2} , \nonumber
\end{eqnarray} 
To simplify the notation we omit the time dependence of $\left<\chi_{1k}(E)\right>$. 
At time $T_R$ we apply the second $\pi/2$ pulse of the Ramsey interferometer (modeled as in section \ref{sec_Model_density_op}), thus the one atom density operator becomes:
\begin{eqnarray}
\widehat{\rho}(T_R) & = & \frac{1}{2} \int dE \frac{E^2}{2}e^{-E} \left|\phi_E\right> \left\{  \left|\uparrow\right> \left[ 1+\left<\chi_{1y}(E)\right> \right] \left<\uparrow\right| \right. \nonumber\\
& & \left. + \left|\downarrow\right> \left[ 1-\left<\chi_{1y}(E)\right> \right] \left<\downarrow\right| \right. \nonumber\\
& & \left. + \left|\uparrow\right> \left[ \left<\chi_{1x}(E)\right>+i\left<\chi_{1z}(E)\right> \right] \left<\downarrow\right| \right.  \nonumber\\ 
& & \left. + \left|\downarrow\right> \left[ \left<\chi_{1x}(E)\right>-i\left<\chi_{1z}(E)\right> \right] \left<\uparrow\right| \right\}  \left<\phi_E\right| .
\end{eqnarray}
The output of the interferometer, i.e. the populations in states $\left|\uparrow\right>$ and $\left|\downarrow\right>$ are:
\begin{eqnarray}
P_\uparrow(T_R) & = & \frac{1}{2}\left[ 1 + \int dE \frac{E^2}{2}e^{-E}\left<\chi_{1y}(E)\right>  \right] \nonumber\\
& = & \frac{1}{2}\left[ 1 + C(T_R)\cos\left( \varphi(T_R) \right)  \right] , \\
P_\downarrow(T_R) & = & \frac{1}{2}\left[ 1 - \int dE \frac{E^2}{2}e^{-E}\left<\chi_{1y}(E)\right>  \right] \nonumber\\
& = & \frac{1}{2}\left[ 1 - C(T_R)\cos\left( \varphi(T_R) \right)  \right] ,
\end{eqnarray}
where the contrast $C$ is defined as:
\begin{eqnarray}
C(T_R) = \left| \int dE \frac{E^2}{2}e^{-E}\left[ \left<\chi_{1y}(E)\right>+i\left<\chi_{1x}(E)\right> \right] \right| ,
\label{eq_Contrast_def}
\end{eqnarray}
and the phase-shift $\varphi$ as:
\begin{eqnarray}
 & & \varphi(T_R) = \nonumber\\
 & & \arg\left\{ \int dE \frac{E^2}{2}e^{-E}  \left[ \left<\chi_{1y}(E)\right>+i\left<\chi_{1x}(E)\right> \right] \right\} .
 \label{eq_PhaseShift_def}
\end{eqnarray}
A $\pi$ phase due to the two consecutive $\pi/2$ pulses has been discarded. 
To define the contrast and the phase-shift, we used the analytic function $\left<\chi_{1y}(E)\right> + i \mathcal{H} [ \left<\chi_{1y}(E)\right> ]$ associated with the function $\left<\chi_{1y}(E)\right>$ where $\mathcal{H}[f]$ is the Hilbert transform \cite{Bendat2011,Zweig1990} of the function $f$. 
Since $\left<\chi_{1y}(E)\right>$ is the quadrature of $\left<\chi_{1x}(E)\right>$, we assume that $\mathcal{H}[\left<\chi_{1y}(E)\right>] = \left<\chi_{1x}(E)\right>$ \cite{Zweig1990,Larkin1996}.

\subsection{Case without interaction}

We first consider the case without interactions, i.e. the case $(4\pi\hbar aN)/(mV_{eff}) \rightarrow 0$  and $1/\tau_{th} \rightarrow 0$. In this case, we can give analytic expressions for the contrast and the phase-shift. The evolution of the one atom average spin is: $\left<\chi_{1x}(E)\right>=\sin\left[ \Omega(E)t \right]$, $\left<\chi_{1y}(E)\right>=-\cos\left[ \Omega(E)t \right]$ and $\left<\chi_{1z}(E)\right>=0$, to be consistent with our model of a $\pi/2$ pulse we took the initial condition: $(\left<\chi_{1x}(E)\right>,\left<\chi_{1y}(E)\right>,\left<\chi_{1z}(E)\right>)=(0,-1,0)$. 
Since $\Omega(E) = E/t_c$, the components of the one atom average spin can be easily integrated over the energy:
\begin{eqnarray}
\left<\chi_{1x}\right> & = & \int dE \frac{E^2}{2}e^{-E}\left<\chi_{1x}(E)\right> \nonumber\\
& = & - \frac{u^3-3u}{u^6+3u^4+3u^2+1} , \nonumber \\
\left<\chi_{1y}\right> & = & \int dE \frac{E^2}{2}e^{-E}\left<\chi_{1y}(E)\right> \nonumber\\
& = & + \frac{3u^2-1}{u^6+3u^4+3u^2+1} , \nonumber\\
\left<\chi_{1z}\right> & = & 0 ,
\label{eq_StotWithoutInt}
\end{eqnarray}
with $u=t/t_c$. Definitions (\ref{eq_Contrast_def}) and (\ref{eq_PhaseShift_def}) give the contrast and the phase-shift:
\begin{eqnarray}
C(u) & = & \frac{1}{\sqrt{(u^2+1)^3}} , \nonumber\\
\varphi(u) & = & - \arctan\left[ (u^3-3u)/(3u^2-1) \right] .
\end{eqnarray}

These results are displayed in figures \ref{fig_ISRE_case_1} and \ref{fig_ISRE_case_2} as solid black curves. 
We recover the result of reference \cite{DupontNivet2014}: in absence of interaction, $t_c$ is the characteristic contrast decay time. This decay is also consistent with the experimental data \cite{DupontNivet2017b}. 
Here, since we neglected interactions, the phase-shift only comes from the potential asymmetry.
From the equation for the phase-shift, one can derive the levels of stability of the gas temperature $T$ and of the potential asymmetry $\omega/\delta\omega$ which are required to achieve a desired phase-shift stability. 

\subsection{Contrast and phase-shift without spatial separation}

We now consider the effect of atom-atom interactions.
First we will examine the interference contrast in the case without spatial separation.
This is the situation in which the identical spin rotation effect can be present. 
We numerically integrate equation (\ref{eq_FullSpinDynamics}) and use the contrast definition (\ref{eq_Contrast_def}). For consistency with previous sections, the initial condition is taken as: $(\left<\chi_{1x}(E)\right>,\left<\chi_{1y}(E)\right>,\left<\chi_{1z}(E)\right>)=(0,-1,0)$. 
We do this for the three interaction kernels computed in section \ref{sec_Kernel}. 
In the three plots displayed in figures \ref{fig_ISRE_case_1}.a, \ref{fig_ISRE_case_1}.b and \ref{fig_ISRE_case_1}.c, $\Omega(E)$ is unchanged and only the atomic density $\overline{n}=N/V_{eff}$ changes. 
To use numbers comparable to reference \cite{Deutsch2010}, we maintain a ratio of 0.27 between $1/\tau_{th}$ and $2\hbar a N/(mV_{eff})$. 

Comparing the curves for $\overline{n}=0$ with those for $\overline{n} > 0$, one sees a slowdown in the contrast decay due to the identical spin rotation effect.
The first term of the right hand side of equation (\ref{eq_FullSpinDynamics}) shows that hot atoms (those with a higher energy) rotate faster in the equatorial plane of the Bloch sphere than cold atoms because they see a larger trap asymmetry ($\Omega(E) \propto E$). 
This leads to dephasing. 
As explained in \cite{Deutsch2010} re-phasing arises from the second term of the right hand side of equation (\ref{eq_FullSpinDynamics}) which describes a rotation of the one atom average spin around its mean value over the energy. 
When the re-phasing term corresponds to a $\pi$ rotation of $\left< \vec{\chi}_1(E) \right>$ around its mean value, 
hot atoms lag behind the cold ones in the equatorial plane of the Bloch sphere, but since they rotate faster they catch up.
This yields a re-phasing and a contrast revival as shown in figure \ref{fig_ISRE_case_1}.
However for the three geometries the behaviour of the contrast decay is different. 
When the interaction kernel becomes less and less long-range in energy (when $K$ decreases faster with $|E-E'|$), the contrast decreases faster. 

The identical spin rotation effect dominates the contrast decay behaviour if the first and the third terms of equation (\ref{eq_FullSpinDynamics}) are smaller than the second term.
In the case of plane waves, we recover two conditions given in reference \cite{Deutsch2010}. 
The first condition is $1/t_c \ll 4\pi\hbar a N/(mV_{eff})$, {\it i.e.} the dephasing must be slower than the rephasing. 
The second condition is $1/\tau_{th} \ll 4\pi\hbar a N/(mV_{eff})$, {\it i.e.} the thermal relaxation of the gas must be slower than the re-phasing mechanism which is equivalent to the condition $a \ll \lambda_{th}$. 

Using the definition (\ref{eq_PhaseShift_def}), we also compute the phase-shift of the Ramsey interferometer.
For the same parameters as in figures \ref{fig_ISRE_case_1}.a, \ref{fig_ISRE_case_1}.b and \ref{fig_ISRE_case_1}.c, we display the phase-shift in figures \ref{fig_ISRE_case_1}.d, \ref{fig_ISRE_case_1}.e and \ref{fig_ISRE_case_1}.f.
The phase-shift is not displayed when the contrast is below 1~\% because for low contrast the numerical computation of the phase-shift becomes less and less relevant.
Even in this case where the three interaction lengths are equal, we clearly see a variation of the slope of the phase-shift curve with the atomic density \cite{Gibble2009,Maineult2012}. 
From this slope variation, one can derive the required level of stability of the gas density $n$ to achieve a desired phase-shift stability. 

\subsection{Contrast and phase-shift with spatial separation: enhancing the interferometer contrast}

\subsubsection{Perfect $\pi/2$ pulse}

In the case of spatial separation of the two internal states, the identical spin rotation effect is absent. 
The contrast and the phase-shift are displayed in figure \ref{fig_ISRE_case_2}, using the same parameters and initial conditions as in figure \ref{fig_ISRE_case_1}. Changing the interaction geometry does not change the evolution of the contrast. 
Changing the atomic density (over the range studied here) does not  significantly change the contrast either but affects the phase-shift. 
In our case this is due to: i)~$\tau_{th} \gg t_c$, i.e. the damping term remains close to zero during the evolution, and to ii)~the hypothesis of perfect $\pi/2$ pulse at the beginning of the Ramsey sequence, i.e. the component of the one-atom average spin along $z$ axis is zero, thus the identical spin rotation effect term of equation (\ref{eq_SplittingSpinDynamics}) also remains close to zero.
This is the case studied in reference \cite{DupontNivet2014}. 

\subsubsection{Imperfect $\pi/2$ pulse}

In equation (\ref{eq_SplittingSpinDynamics}) the one atom average spin dephasing arises from $\Omega(E)$, which is related to the difference in the two potentials. 
In the following we will see that this effect can be reduced with a proper choice of $\left<\chi_{1z}(E)\right>$. 
A non zero $\left<\chi_{1z}(E)\right>$ is created by an imperfect $\pi/2$ pulse, either because of a detuning from resonance or by an imperfect pulse duration.  
To model these two defects we define two dimensionless parameters $\alpha$ and $\delta'$:
\begin{eqnarray}
\Omega_R t = \frac{\pi}{2}\left(1+\alpha \right) , \qquad
\delta' = \frac{\delta}{\Omega_R} ,
\end{eqnarray}
where $\Omega_R$ is the Rabi frequency and $\delta$ the detuning.
 The parameter $\alpha$ thus describes a variation in the pulse duration and $\delta'$ is a normalized detuning. 
Up to first order in $\delta'$ and $\alpha$, the effect of an imperfect $\pi/2$ pulse is given by:
\begin{eqnarray}
\hspace{-0.3cm} \left|\uparrow\right> & \rightarrow & \frac{1}{\sqrt{2}} \left[ \left( 1 - \frac{\pi}{2}\alpha -i \delta' \right) \left|\uparrow\right> -i\left( 1 + \frac{\pi}{2}\alpha \right) \left|\downarrow\right> \right] , \nonumber\\
\hspace{-0.3cm} \left|\downarrow\right> & \rightarrow & \frac{1}{\sqrt{2}} \left[ -i\left( 1 + \frac{\pi}{2}\alpha \right) \left|\uparrow\right> + \left( 1 - \frac{\pi}{2}\alpha + i\delta' \right) \left|\downarrow\right> \right] 
\end{eqnarray}
Following the same procedure as in the beginning of section \ref{sec_Numeric}, we compute the populations at the output of the Ramsey interferometer (the change in the model of the $\pi/2$ pulses does not change the equations (\ref{eq_FullSpinDynamics}) and (\ref{eq_SplittingSpinDynamics}) governing the one atom average spin):
\begin{eqnarray}
P_\uparrow(T_R) & = & \frac{1}{2} \left[ 1 - \frac{\pi}{2}\alpha\int dE \frac{E^2}{2}e^{-E}\left<\chi_{1z}(E)\right> \right. \nonumber\\ 
& & \hspace{2cm} \left. + C(T_R)\cos\left( \varphi(T_R) \right)  \right] , \nonumber\\
P_\downarrow(T_R) & = & \frac{1}{2} \left[ 1 + \frac{\pi}{2}\alpha\int dE \frac{E^2}{2}e^{-E}\left<\chi_{1z}(E)\right> \right. \nonumber\\ 
& & \hspace{2cm} \left. - C(T_R)\cos\left( \varphi(T_R) \right)  \right] .
\end{eqnarray}
The contrast is defined as $C(t)=\left|A(t)\right|$ and the phase-shift as $\varphi(t)=\arg\left[ A(t) \right]$, with:
\begin{eqnarray}
A(t) & = & \int dE \frac{E^2}{2} e^{-E} \left\{ \left<\chi_{1y}(E)\right> + \delta'\left<\chi_{1x}(E)\right> \right. \nonumber\\
& & \hspace{1.3cm} \left. +i \left[ \left<\chi_{1x}(E)\right> - \delta'\left<\chi_{1y}(E)\right>  \right] \right\} .
\end{eqnarray}
Note that the term $\frac{\pi}{2}\alpha\int dE \frac{E^2}{2}e^{-E}\left<\chi_{1z}(E)\right>$ appearing in the population is not included in the contrast definition because it simply shifts the center of the fringes.

The normalized detuning $\delta'$ takes into account the variation of the energy difference $E_\uparrow(n) - E_\downarrow(n)$ with $n$ (the same notation and reasoning as in the calculation of $\Omega(E)$ in section \ref{sec_Introduction} are used). 
Using the hypothesis that the $\pi/2$ pulse is tuned to be resonant with the transition linking the two ground states of the two trapping potentials, one can show that $\delta'(E) = E/(\Omega_R t_c)$.

Using $\delta'(E) = E/(\Omega_R t_c)$, we show a numerical simulation of the contrast and the phase-shift in figures \ref{fig_MeanField} and \ref{fig_ISRE_case_2_enhanced} in the case of an isotropic 3D trapping potential (interaction kernel of equation (\ref{eq_3DKernel})) and two different values of $\alpha$.
The initial condition is taken as: $(\left<\chi_{1x}(E)\right>,\left<\chi_{1y}(E)\right>,\left<\chi_{1z}(E)\right>)=(0,-\sqrt{1-(\Delta N)^2},\Delta N)$, with $\Delta N=P_\uparrow-P_\downarrow=-\pi \alpha/2$ the population imbalance after the first $\pi/2$ pulse (it is reminded that in equation (\ref{eq_DensityOp}) before the first $\pi/2$ pulse all the atoms are in the state $\left|\uparrow\right>$). 

The case without asymmetry between the two trapping potentials ($\Omega(E)=0$) is shown in figure \ref{fig_MeanField}. 
Whatever the sign of $\alpha$, the contrast decays in the same way.  
Changing the sign of $\alpha$ changes the sign of the population imbalance and thus the sign of the phase-shift as can be seen from the curves. 

In the presence of an asymmetry between the two trapping potentials the behavior is different. 
If $\alpha=0$ (figures \ref{fig_ISRE_case_2_enhanced}.b and \ref{fig_ISRE_case_2_enhanced}.e) we find the same behaviour as in figure \ref{fig_ISRE_case_2}, meaning that the energy variation of the detuning $\delta'(E) = E/(\Omega_R t_c)$ plays no role. 
This is due to our choice of $\Omega_R=2\pi\times500$~Hz and $T_R \leq 3$~s which renders   
$\delta'(E)/(\Omega(E)T_R)$ negligible. 
If $\alpha>0$ the contrast decreases faster than in the case $\alpha=0$ (figure \ref{fig_ISRE_case_2_enhanced}.c). 
On the other hand if $\alpha<0$ the decay can be slowed down. 
As shown in figure \ref{fig_ISRE_case_2_enhanced}.a a density corresponding to $2\hbar a N /(mV_{eff})\simeq$~7~Hz can increase the decay time by nearly an order of magnitude for $\alpha=0.15$. 
This happens because in this situation the mean field shift and the effect of the trap asymmetry have opposite signs.


\section{Conclusion}
\label{sec_conclusion}

We have derived an equation of evolution for the one atom average spin in the presence of atom-atom interactions. This equation allowed us to compute the contrast and the phase-shift of a trapped Ramsey interferometer with internal state labelling. 
We have computed these two quantities with and without spatial separation of the two arms of the interferometer. 
In the case without splitting, the trapping geometry plays an important role for the damping time of the contrast, via the identical spin rotation effect.
When the clouds are split, the ISRE is absent, but interaction effects can still increase the contrast decay time, because mean field shifts can partly offset the effect of any residual trap asymmetry.
Since the interactions significantly contribute to the overall phase of the interferometer, this work also highlights the importance of controlling the atomic density for future applications of trapped atom interferometers.

\vspace*{0.3cm}
\begin{acknowledgments}
This work has been carried out within the OnACIS project ANR-13-ASTR-0031 and the NIARCOS project ANR-18-ASMA-0007-02 funded by the French National Research Agency (ANR) in the frame of its 2013 Astrid and 2018 Astrid Maturation programs. S. Schwartz acknowledges funding from the European Union under the Marie Sklodowska Curie Individual Fellowship Programme H2020-MSCA-IF-2014 (project number 658253).
\end{acknowledgments}


\appendix

\section{Demonstration of the equation for the one atom average spin}
\label{sec_Demonstration}

Before starting the commutator calculus of equation (\ref{eq_SpinOp}), as a first step we need to express $\left<\vec{S}_1(E)\right>$ in terms of the Pauli matrices.
First we compute the product of the one atom density operator and the one atom spin operator, then using definition (\ref{eq_OneAtomAverage}) we take the average:
\begin{eqnarray}
\left<S_{1x}(E)\right> & = & \frac{1}{4} e^{-E} \left( \sigma_x + i\sigma_z \right) ,  \nonumber\\
\left<S_{1y}(E)\right> & = & \frac{1}{4} e^{-E} \left( \sigma_y - Id \right) ,  \nonumber\\
\left<S_{1z}(E)\right> & = & \frac{1}{4} e^{-E} \left( \sigma_z - i\sigma_x \right) .
\label{eq_OneAtomAverageSpinComp}
\end{eqnarray}

\subsection{First term of the Hamiltonian}

The first term of the Hamiltonian $\widehat{H}^{mean}_0$ commutes with the one atom spin operator, thus this term contributes to zero in the one atom average spin equation.

\subsection{Second term of the Hamiltonian: potential asymmetry}

First we compute the commutator between $\widehat{H}_0^{diff}$ and $\vec{S}_1$, then the products of the last commutators with the one atom density operator gives:
\begin{eqnarray}
\widehat{\rho}\left[\widehat{H}_0^{diff},S_{1x}\right] & = & +\frac{i}{4} \sum_E e^{-E} \hbar\Omega(E) \left|\phi_E\right>\left( \sigma_y - Id \right) \left<\phi_E\right| , \nonumber\\
\widehat{\rho}\left[\widehat{H}_0^{diff},S_{1y}\right] & = & -\frac{i}{4} \sum_E e^{-E} \hbar\Omega(E) \left|\phi_E\right>\left( \sigma_x + i\sigma_z \right) \left<\phi_E\right| ,  \nonumber\\
\widehat{\rho}\left[\widehat{H}_0^{diff},S_{1z}\right] & = & 0 .
\end{eqnarray}
Finally, using definition of the one atom average (\ref{eq_OneAtomAverage}) and identifying the components of the one atom average spin (\ref{eq_OneAtomAverageSpinComp}), we obtain the trace:
\begin{eqnarray}
\left< \left[\widehat{H}_0^{diff},\vec{S}_{1}\right] \right> & = & \left| \begin{matrix}
+ i\hbar\Omega(E) \left<S_{1y}(E)\right> \\
- i\hbar\Omega(E) \left<S_{1x}(E)\right> \\
0
\end{matrix} \right. \nonumber\\
& = & -i\hbar \left| \begin{matrix}
0 \\
0 \\
\Omega(E)
\end{matrix} \right. \wedge \left<\vec{S}_{1}(E)\right> .
\label{eq_Asym_Term}
\end{eqnarray}

\subsection{Third term of the Hamiltonian: interactions}

To compute the effect of the interaction Hamiltonian, the two atom density operator $\widehat{\rho}_{12}$ is needed. We define it as the tensor product of two one atom density operators:
\begin{eqnarray}
\widehat{\rho}_{12} & = & \frac{1}{4} \sum_{E_a,E_b}  e^{-E_a} e^{-E_b} \left|\phi_{E_a}\right>_1 \left(Id^1-\sigma_y^1 \right) {}_{1} \! \left<\phi_{E_a}\right| \nonumber\\
& & \hspace{2cm} \otimes \left|\phi_{E_b}\right>_2 \left(Id^2-\sigma_y^2 \right) {}_{2} \! \left<\phi_{E_b}\right| ,
\end{eqnarray}
and we define the two atom average as:
\begin{eqnarray}
\left<\widehat{X}(E)\right> & = & \sum_{E',E''} {}_{1} \! \left<\phi_{E'}\right| {}_{2} \! \left<\phi_{E''}\right| \nonumber\\
& & \left( \left|\phi_E\right>_1 {}_{1} \! \left<\phi_E\right| \widehat{\rho}_{12} \widehat{X} \right) \left|\phi_{E'}\right>_1 \left|\phi_{E''}\right>_2 .
\label{eq_TwoAtomAverage}
\end{eqnarray}

\subsubsection{$\left[\widehat{H}_{int},S_{1x}\right]$ calculation}  

Let's start with the commutator between the interaction Hamiltonian and the $x$ component of the one atom spin operator. In the case $a_{\uparrow\downarrow}=a_{\uparrow\uparrow}=a_{\downarrow\downarrow}=a$ (for example a trapped atomic clock using rubidium 87), after summing all the six terms given in appendix \ref{sec_HintSx}:
\begin{eqnarray}
& & \widehat{\rho}_{12}\left[ \widehat{H}_{int},S_{1x}\right] = \nonumber\\
& & \qquad i \frac{4\pi\hbar^2a}{m} \frac{N}{16} \sum_{E_a,E_b} e^{-E_a} e^{-E_b} \sum_{E_3} I^{E_a,E_b}_{E_a-E_3,E_b+E_3}   \nonumber\\
& & \qquad \left\{ \left|\phi_{E_a}\right>_1 \left( \sigma_z^1-i\sigma_x^1 \right) {}_{1} \! \left<\phi_{E_a-E_3}\right| \right. \nonumber\\
& & \qquad \qquad \qquad \left. \otimes \left|\phi_{E_b}\right>_2 \left( Id^2-\sigma_y^2 \right) {}_{2} \! \left<\phi_{E_b+E_3}\right| \right.  \nonumber\\
& & \qquad \left. - \left|\phi_{E_a}\right>_1 \left( Id^1-\sigma_y^1 \right) {}_{1} \! \left<\phi_{E_a-E_3}\right| \right. \nonumber\\
& & \qquad \qquad \qquad \left. \otimes \left|\phi_{E_b}\right>_2 \left( \sigma_z^2-i\sigma_x^2 \right) {}_{2} \! \left<\phi_{E_b+E_3}\right| \right\} .
\end{eqnarray}
From the last equation, using the definition (\ref{eq_TwoAtomAverage}), identifying the components of the one atom average spin (equation (\ref{eq_OneAtomAverageSpinComp})) and changing the notation $\sum_E$ for $\int (E^2/2) dE$, we compute the trace:
\begin{eqnarray}
& & \hspace{-0.5cm} \left< \left[\widehat{H}_{int},S_{1x}\right] \right> = - i \frac{4\pi\hbar^2aN}{m} \int dE' \frac{E'^2}{2} I^{E,E'}_{E,E'}  \nonumber\\
& & \times \left\{ \left<S_{1z}(E)\right> \left<S_{1y}(E')\right>  - \left<S_{1y}(E)\right> \left<S_{1z}(E')\right>  \right\} .
\label{eq_EffetIntSx}
\end{eqnarray}

\subsubsection{$\left[\widehat{H}_{int},S_{1y}\right]$ calculation} 

We continue with the $y$ component of the one atom spin. In the case $a_{\uparrow\downarrow}=a_{\uparrow\uparrow}=a_{\downarrow\downarrow}=a$, after summing all the six terms given in appendix \ref{sec_HintSy}:
\begin{eqnarray}
& & \widehat{\rho}_{12}\left[ \widehat{H}_{int},S_{1y}\right] = \nonumber\\
& & \qquad i \frac{4\pi\hbar^2a}{m} \frac{N}{16} \sum_{E_a,E_b} e^{-E_a} e^{-E_b} \sum_{E_3} I^{E_a,E_b}_{E_a-E_3,E_b+E_3}   \nonumber\\
& & \qquad \left\{ \left|\phi_{E_a}\right>_1 \left( \sigma_z^1-i\sigma_x^1 \right) {}_{1} \! \left<\phi_{E_a-E_3}\right| \right. \nonumber\\
& & \qquad \qquad \qquad \left. \otimes \left|\phi_{E_b}\right>_2 \left( \sigma_x^2+i\sigma_z^2 \right) {}_{2} \! \left<\phi_{E_b+E_3}\right| \right.  \nonumber\\
& & \qquad \left. - \left|\phi_{E_a}\right>_1 \left( \sigma_x^1+i\sigma_z^1 \right) {}_{1} \! \left<\phi_{E_a-E_3}\right| \right. \nonumber\\
& & \qquad \qquad \qquad \left.  \otimes \left|\phi_{E_b}\right>_2 \left( \sigma_z^2-i\sigma_x^2 \right) {}_{2} \! \left<\phi_{E_b+E_3}\right| \right\} .
\end{eqnarray}
From the last equation we compute the trace:
\begin{eqnarray}
& & \hspace{-0.5cm} \left< \left[\widehat{H}_{int},S_{1y}\right] \right> = - i \frac{4\pi\hbar^2aN}{m} \int dE' \frac{E'^2}{2} I^{E,E'}_{E,E'}  \nonumber\\
& & \times \left\{ \left<S_{1x}(E)\right> \left<S_{1z}(E')\right>  - \left<S_{1z}(E)\right> \left<S_{1x}(E')\right>  \right\} .
\label{eq_EffetIntSy}
\end{eqnarray}

\subsubsection{$\left[\widehat{H}_{int},S_{1z}\right]$ calculation} 

Finally we calculate the $z$ component of the one atom spin operator. In the case $a_{\uparrow\downarrow}=a_{\uparrow\uparrow}=a_{\downarrow\downarrow}=a$, after summing all the six terms given in appendix \ref{sec_HintSz}:
\begin{eqnarray}
& & \widehat{\rho}_{12}\left[ \widehat{H}_{int},S_{1z}\right] = \nonumber\\
& & \qquad i \frac{4\pi\hbar^2a}{m} \frac{N}{16} \sum_{E_a,E_b} e^{-E_a} e^{-E_b} \sum_{E_3} I^{E_a,E_b}_{E_a-E_3,E_b+E_3}   \nonumber\\
& & \qquad \left\{ \left|\phi_{E_a}\right>_1 \left( Id^2-\sigma_y^1 \right) {}_{1} \! \left<\phi_{E_a-E_3}\right| \right. \nonumber\\
& & \qquad \qquad \qquad \left. \otimes \left|\phi_{E_b}\right>_2 \left( \sigma_x^2+i\sigma_z^2 \right) {}_{2} \! \left<\phi_{E_b+E_3}\right| \right.  \nonumber\\
& & \qquad \left. - \left|\phi_{E_a}\right>_1 \left( \sigma_x^2+i\sigma_z^2 \right) {}_{1} \! \left<\phi_{E_a-E_3}\right| \right. \nonumber\\
& & \qquad \qquad \qquad \left. \otimes \left|\phi_{E_b}\right>_2 \left( Id^2-\sigma_y^2 \right) {}_{2} \! \left<\phi_{E_b+E_3}\right| \right\} .
\end{eqnarray}
From the last equation we compute the trace:
\begin{eqnarray}
& & \hspace{-0.5cm} \left< \left[\widehat{H}_{int},S_{1z}\right] \right> = - i \frac{4\pi\hbar^2aN}{m} \int dE' \frac{E'^2}{2} I^{E,E'}_{E,E'}  \nonumber\\
& \times & \left\{ \left<S_{1y}(E)\right> \left<S_{1x}(E')\right>  - \left<S_{1x}(E)\right> \left<S_{1y}(E')\right>  \right\} .
\label{eq_EffetIntSz}
\end{eqnarray}

In conclusion, summing up the results of the last three paragraphs, we have:
\begin{eqnarray}
& & \left< \left[\widehat{H}_{int},\vec{S}_{1}\right] \right> = - i\frac{4\pi\hbar^2aN}{m} \nonumber\\
& \times & \int dE' \frac{E'^2}{2} I^{E,E'}_{E,E'}  \left<\vec{S}_{1}(E')\right> \wedge \left<\vec{S}_{1}(E)\right> .
\label{eq_ISRE_Term}
\end{eqnarray}

\subsection{Summing up the previous results}
\label{sec_DynamicEqWithoutDecay}

From equation (\ref{eq_Spin_Average}), (\ref{eq_Asym_Term}) and (\ref{eq_ISRE_Term}) we deduce a dynamical equation describing the average spin of a trial atom of energy $E$ in a gas of N identical atoms:
\begin{eqnarray}
& & \frac{d}{dt} \left<\vec{S}_1(E)\right> \equiv \left<\frac{d}{dt}\vec{S}_1(E)\right> =  \left| \begin{matrix}
0 \\
0 \\
\Omega(E) \end{matrix} \right. \wedge \left<\vec{S}_{1}(E)\right> \nonumber\\
& & + \frac{4\pi\hbar aN}{m} \int dE' \frac{E'^2}{2} I^{E,E'}_{E,E'} \left<\vec{S}_{1}(E')\right> \wedge \left<\vec{S}_{1}(E)\right> \nonumber\\
& & + \left< \frac{\partial}{\partial t} \vec{S}_1(E) \right> . 
\label{eq_AveSpinWithoutDampingOld}
\end{eqnarray}
To write equation (\ref{eq_AveSpinWithoutDampingOld}) and initial condition with the convention of reference \cite{Deutsch2010} the following change of notation is used: $\left<\vec{S}_{1}(E)\right>=e^{-E}\left<\vec{\chi}_{1}(E)\right>$:
\begin{eqnarray}
& & \frac{d}{dt} \left<\vec{\chi}_1(E)\right> =  +  \left| \begin{matrix}
0 \\
0 \\
\Omega(E) \end{matrix} \right. \wedge \left<\vec{\chi}_{1}(E)\right> \nonumber\\
& & \hspace{-0.5cm} + \frac{4\pi\hbar aN}{m} \int dE' \frac{E'^2}{2} e^{-E'} I^{E,E'}_{E,E'} \left<\vec{\chi}_{1}(E')\right> \wedge \left<\vec{\chi}_{1}(E)\right> \nonumber\\
& & \hspace{-0.5cm} + \left< \frac{\partial}{\partial t} \vec{\chi}_1(E) \right> .
\label{eq_AveSpinWithoutDamping}
\end{eqnarray}
The third term (spin decay) is introduced in the following section.

\subsection{Introducing spin decay}

The damping term in the one atom average spin equation comes from two-body collisions. 
These collisions re-thermalize the gas and change the energies of the colliding atoms and thus force the energy distribution of the one atom average spin $\left<\vec{\chi}_{1}(E)\right>$ to stay close to its equilibrium value which is given by $\int dE \frac{E^2}{2}e^{-E} \left<\vec{\chi}_{1}(E)\right> $ (see the next paragraph). This will limit the identical spin rotation effect. Using the same hypothesis as in section \ref{sec_Introduction} ($\lambda_{th} \gg a_{ij}$) to reduce the interaction Hamiltonian to s-wave scattering, we suppose that the gas is in a regime where the interaction described by the Hamiltonian (\ref{eq_HInt}) dominates the previously mentioned collisions \cite{Walraven2010}. The same hypothesis is necessary to observe spin waves in a gas \cite{Laloe1988,Bigelow1989}. Thus to take these collisions into account we approximate the collision integral by a relaxation term \cite{Reif1965} and we identify it with the last term of the right hand side of equation (\ref{eq_AveSpinWithoutDamping}): 
\begin{eqnarray}
& & \left< \frac{\partial}{\partial t} \vec{\chi}_1(E) \right> = \nonumber\\
& & - \frac{1}{\tau_{th}} \left( \left<\vec{\chi}_{1}(E)\right> - \int dE \frac{E^2}{2}e^{-E} \left<\vec{\chi}_{1}(E)\right>  \right) ,
\label{eq_SpinDamping}
\end{eqnarray}
where $\tau_{th}$ is the thermal relaxation time. This term forces the average spin $\left<\vec{\chi}_{1}(E)\right>$ to fit to its equilibrium value. The thermalization time is linked to the collision time $\tau_c$: $\tau_{th}^{-1}=\tau_c^{-1}/3$ \cite{Walraven2010}, where the collision time is $\tau_c =\overline{n}v_r\sigma$, with $\overline{n}$ the gas density, $v_r = \sqrt{16k_BT/(\pi m)}$ the mean relative thermal velocity and $\sigma=8\pi a^2$ the two-body collision cross section for bosons \cite{Walraven2010}. To go beyond this relaxation time approximation, the reader is referred to \cite{Bradley2002,Gardiner1997}.

\subsection{Equilibrium value of $\left<\vec{\chi}_1(E)\right>$}
\label{sec_AppendixEqui}

To derive the equilibrium value of $\left<\vec{\chi}_1(E)\right>$, we need to go back to the position $r$ and the momentum $p$ description of the $\left<\vec{S}_1\right>$. 
At equilibrium: $\left<\vec{S}_1(r,p,t)\right>=(\exp(-p^2/2)/\sqrt{2\pi})\int dp'^3 \left<\vec{S}_1(r,p',t)\right>$ \cite{Fuchs2002,Solaro2016b}, which can be written in angle $\alpha$ and energy $E$ variables: $\left<\vec{S}_1(\alpha,E,t)\right>=\exp(-E)\int dE' (E'^2/2) \left<\vec{S}_1(\alpha,E',t)\right>$. In the Knudsen regime, we can integrate over the angle variable $\alpha$ thus: $\left<\vec{S}_1(E,t)\right>=\exp(-E)\int dE' (E'^2/2) \left<\vec{S}_1(E',t)\right>$. Finally performing the change of notation $\left<\vec{S}_1(E)\right> = \exp(-E) \left<\vec{\chi}_1(E)\right>$ used in appendix \ref{sec_DynamicEqWithoutDecay}, the equilibrium value of $\left<\vec{\chi}_1(E)\right>$ is $\int dE \exp(-E) (E^2/2) \left<\vec{\chi}_1(E)\right>$.

\subsection{Case with spatial separation}
\label{sec_AppendixSplit}

In the case with spatial separation, to reuse the previous results we suppose that $a_{\uparrow\downarrow}=0$ and $a_{\uparrow\uparrow}=a_{\downarrow\downarrow}=a$ (we consider rubidium 87). More precisely, $a_{\uparrow\downarrow}$ is not zero but we suppose that the spatial separation between the two spin states is enough to neglect the wave function overlap (\ref{eq_WaveOverLap}). Thus equations (\ref{eq_EffetIntSx}), (\ref{eq_EffetIntSy}) and (\ref{eq_EffetIntSz}) become:
\begin{eqnarray}
\left< \left[\widehat{H}_{int},S_{1x}\right] \right> & = & + i \frac{8\pi\hbar^2aN}{m} \nonumber\\
& & \hspace{-1.2cm} \times \int dE' \frac{E'^2}{2} I^{E,E'}_{E,E'}  \left<S_{1y}(E)\right> \left<S_{1z}(E')\right> ,  \nonumber
\end{eqnarray}

\begin{eqnarray}
\left< \left[\widehat{H}_{int},S_{1y}\right] \right> & = & - i \frac{8\pi\hbar^2aN}{m} \nonumber\\
& & \hspace{-1.2cm} \times \int dE' \frac{E'^2}{2} I^{E,E'}_{E,E'}  \left<S_{1x}(E)\right> \left<S_{1z}(E')\right> , \nonumber\\
\left< \left[\widehat{H}_{int},S_{1z}\right] \right> & = & 0 ,
\end{eqnarray}

\section{Effect of the interaction}
\label{sec_Appendix}

In this appendix, we give more details about the calculation of the effect of the interaction on the one atom average spin. The interaction Hamiltonian (\ref{eq_HInt}) is the sum of six terms. For simplicity in the following they are noted $\widehat{H}_{int} ^i$, with $i=\left\{1,...,6\right\}$, in the same order they appear in equation (\ref{eq_HInt}).

\begin{widetext}
\subsection{Commutator $\left[\widehat{H}_{int},S_{1x}\right]$}
\label{sec_HintSx}

The first term of the interaction Hamiltonian gives:
\begin{flalign}
& \left[ \widehat{H}_{int}^1,S_{1x}\right] = \frac{4\pi\hbar^2a_{\uparrow\downarrow}}{m} \frac{N}{4} \sum_{E_1,E_2,E_3} I^{E_1+E_3,E_2-E_3}_{E_1,E_2} \left|\phi_{E_1+E_3}\right>_1 \footnotesize{\begin{pmatrix}
1 & 0 \\
0 & - 1 \end{pmatrix}}_1 {}_{1} \! \left<\phi_{E_1}\right| \otimes \left|\phi_{E_2-E_3}\right>_2 \footnotesize{\begin{pmatrix}
0 & 0 \\
1 & 0 \end{pmatrix}}_2 {}_{2} \! \left<\phi_{E_2}\right| , & \nonumber
\end{flalign}
\begin{flalign}
& \widehat{\rho}_{12}\left[ \widehat{H}_{int}^1,S_{1x}\right] = \frac{4\pi\hbar^2a_{\uparrow\downarrow}}{m} \frac{N}{16} \sum_{\substack{E_a,E_b \\ E_3}} e^{-E_a-E_b} I^{E_a,E_b}_{E_a-E_3,E_b+E_3} \left|\phi_{E_a}\right>_1 \footnotesize{\begin{pmatrix}
1 & -i \\
-i & - 1 \end{pmatrix}}_1 {}_{1} \! \left<\phi_{E_a-E_3}\right| \otimes \left|\phi_{E_b}\right>_2 \footnotesize{\begin{pmatrix}
i & 0 \\
1 & 0 \end{pmatrix}}_2 {}_{2} \! \left<\phi_{E_b+E_3}\right| . & \nonumber
\end{flalign}
The second term of the interaction Hamiltonian gives:
\begin{flalign}
& \left[ \widehat{H}_{int}^2,S_{1x}\right] = \frac{4\pi\hbar^2a_{\uparrow\downarrow}}{m} \frac{N}{4} \sum_{E_1,E_2,E_3} I^{E_1+E_3,E_2-E_3}_{E_1,E_2}  \left|\phi_{E_1+E_3}\right>_1 \footnotesize{\begin{pmatrix}
-1 & 0 \\
0 &  1 \end{pmatrix}}_1 {}_{1} \! \left<\phi_{E_1}\right| \otimes \left|\phi_{E_2-E_3}\right>_2 \footnotesize{\begin{pmatrix}
0 & 1 \\
0 & 0 \end{pmatrix}}_2 {}_{2} \! \left<\phi_{E_2}\right| , & \nonumber
\end{flalign}
\begin{flalign}
& \widehat{\rho}_{12}\left[ \widehat{H}_{int}^2,S_{1x}\right] = \frac{4\pi\hbar^2a_{\uparrow\downarrow}}{m} \frac{N}{16} \sum_{\substack{E_a,E_b \\ E_3}} e^{-E_a-E_b} I^{E_a,E_b}_{E_a-E_3,E_b+E_3}  \left|\phi_{E_a}\right>_1 \footnotesize{\begin{pmatrix}
-1 & i \\
i & 1 \end{pmatrix}}_1 {}_{1} \! \left<\phi_{E_a-E_3}\right| \otimes \left|\phi_{E_b}\right>_2 \footnotesize{\begin{pmatrix}
0 & 1 \\
0 & -i \end{pmatrix}}_2 {}_{2} \! \left<\phi_{E_b+E_3}\right| . & \nonumber
\end{flalign}
The third term of the interaction Hamiltonian gives:
\begin{flalign}
& \left[ \widehat{H}_{int}^3,S_{1x}\right] = \frac{4\pi\hbar^2a_{\uparrow\downarrow}}{m} \frac{N}{4} \sum_{E_1,E_2,E_3} I^{E_1+E_3,E_2-E_3}_{E_1,E_2} \left|\phi_{E_1+E_3}\right>_1 \footnotesize{\begin{pmatrix}
0 & 1 \\
-1 & 0 \end{pmatrix}}_1 {}_{1} \! \left<\phi_{E_1}\right| \otimes \left|\phi_{E_2-E_3}\right>_2 \footnotesize{\begin{pmatrix}
0 & 0 \\
0 & 1 \end{pmatrix}}_2 {}_{2} \! \left<\phi_{E_2}\right| , & \nonumber
\end{flalign}
\begin{flalign}
& \widehat{\rho}_{12}\left[ \widehat{H}_{int}^3,S_{1x}\right] = \frac{4\pi\hbar^2a_{\uparrow\downarrow}}{m} \frac{N}{16} \sum_{\substack{E_a,E_b \\ E_3}} e^{-E_a-E_b} I^{E_a,E_b}_{E_a-E_3,E_b+E_3} \left|\phi_{E_a}\right>_1 \footnotesize{\begin{pmatrix}
-i & 1 \\
-1 & -i \end{pmatrix}}_1 {}_{1} \! \left<\phi_{E_a-E_3}\right| \otimes \left|\phi_{E_b}\right>_2 \footnotesize{\begin{pmatrix}
0 & i \\
0 & 1 \end{pmatrix}}_2 {}_{2} \! \left<\phi_{E_b+E_3}\right| . & \nonumber
\end{flalign}
The fourth term of the interaction Hamiltonian gives:
\begin{flalign}
& \left[ \widehat{H}_{int}^4,S_{1x}\right] = \frac{4\pi\hbar^2a_{\uparrow\downarrow}}{m} \frac{N}{4} \sum_{E_1,E_2,E_3} I^{E_1+E_3,E_2-E_3}_{E_1,E_2} \left|\phi_{E_1+E_3}\right>_1 \footnotesize{\begin{pmatrix}
0 & -1 \\
1 & 0 \end{pmatrix}}_1 {}_{1} \! \left<\phi_{E_1}\right| \otimes \left|\phi_{E_2-E_3}\right>_2  \footnotesize{\begin{pmatrix}
1 & 0 \\
0 & 0 \end{pmatrix}}_2 {}_{2} \! \left<\phi_{E_2}\right| , & \nonumber
\end{flalign}
\begin{flalign}
& \widehat{\rho}_{12}\left[ \widehat{H}_{int}^4,S_{1x}\right] = \frac{4\pi\hbar^2a_{\uparrow\downarrow}}{m} \frac{N}{16} \sum_{\substack{E_a,E_b \\ E_3}} e^{-E_a-E_b} I^{E_a,E_b}_{E_a-E_3,E_b+E_3} \left|\phi_{E_a}\right>_1 \footnotesize{\begin{pmatrix}
i & -1 \\
1 & i \end{pmatrix}}_1 {}_{1} \! \left<\phi_{E_a-E_3}\right| \otimes \left|\phi_{E_b}\right>_2  \footnotesize{\begin{pmatrix}
1 & 0 \\
-i & 0 \end{pmatrix}}_2 {}_{2} \! \left<\phi_{E_b+E_3}\right| . & \nonumber
\end{flalign}
The fifth term of the interaction Hamiltonian gives:
\begin{flalign}
& \left[ \widehat{H}_{int}^5,S_{1x}\right] = \frac{4\pi\hbar^2a_{\uparrow\uparrow}}{m} \frac{N}{2} \sum_{E_1,E_2,E_3} I^{E_1+E_3,E_2-E_3}_{E_1,E_2}  \left|\phi_{E_1+E_3}\right>_1 \footnotesize{\begin{pmatrix}
0 & 1 \\
-1 & 0 \end{pmatrix}}_1 {}_{1} \! \left<\phi_{E_1}\right| \otimes \left|\phi_{E_2-E_3}\right>_2  \footnotesize{\begin{pmatrix}
1 & 0 \\
0 & 0 \end{pmatrix}}_2 {}_{2} \! \left<\phi_{E_2}\right| , & \nonumber
\end{flalign}
\begin{flalign}
& \widehat{\rho}_{12}\left[ \widehat{H}_{int}^5,S_{1x}\right] = \frac{4\pi\hbar^2a_{\uparrow\uparrow}}{m} \frac{N}{8} \sum_{\substack{E_a,E_b \\ E_3}} e^{-E_a-E_b} I^{E_a,E_b}_{E_a-E_3,E_b+E_3} \left|\phi_{E_a}\right>_1 \footnotesize{\begin{pmatrix}
-i & 1 \\
-1 & -i \end{pmatrix}}_1 {}_{1} \! \left<\phi_{E_a-E_3}\right| \otimes \left|\phi_{E_b}\right>_2  \footnotesize{\begin{pmatrix}
1 & 0 \\
-i & 0 \end{pmatrix}}_2 {}_{2} \! \left<\phi_{E_b+E_3}\right| . & \nonumber
\end{flalign}
The sixth term of the interaction Hamiltonian gives:
\begin{flalign}
& \left[ \widehat{H}_{int}^6,S_{1x}\right] = \frac{4\pi\hbar^2a_{\downarrow\downarrow}}{m} \frac{N}{2} \sum_{E_1,E_2,E_3} I^{E_1+E_3,E_2-E_3}_{E_1,E_2}  \left|\phi_{E_1+E_3}\right>_1 \footnotesize{\begin{pmatrix}
0 & -1 \\
1 & 0 \end{pmatrix}}_1 {}_{1} \! \left<\phi_{E_1}\right| \otimes \left|\phi_{E_2-E_3}\right>_2  \footnotesize{\begin{pmatrix}
0 & 0 \\
0 & 1 \end{pmatrix}}_2 {}_{2} \! \left<\phi_{E_2}\right| , & \nonumber
\end{flalign}
\begin{flalign}
& \widehat{\rho}_{12}\left[ \widehat{H}_{int}^6,S_{1x}\right] = \frac{4\pi\hbar^2a_{\downarrow\downarrow}}{m} \frac{N}{8} \sum_{\substack{E_a,E_b \\ E_3}} e^{-E_a-E_b} I^{E_a,E_b}_{E_a-E_3,E_b+E_3} \left|\phi_{E_a}\right>_1 \footnotesize{\begin{pmatrix}
i & -1 \\
1 & i \end{pmatrix}}_1 {}_{1} \! \left<\phi_{E_a-E_3}\right| \otimes \left|\phi_{E_b}\right>_2  \footnotesize{\begin{pmatrix}
0 & i \\
0 & 1 \end{pmatrix}}_2 {}_{2} \! \left<\phi_{E_b+E_3}\right| . & \nonumber
\end{flalign}

\subsection{Commutator $\left[\widehat{H}_{int},S_{1y}\right]$}
\label{sec_HintSy}

The first term of the interaction Hamiltonian gives:
\begin{flalign}
& \left[ \widehat{H}_{int}^1,S_{1y}\right] = \frac{4\pi\hbar^2a_{\uparrow\downarrow}}{m} \frac{N}{4} \sum_{E_1,E_2,E_3} I^{E_1+E_3,E_2-E_3}_{E_1,E_2}   \left|\phi_{E_1+E_3}\right>_1 \footnotesize{\begin{pmatrix}
i & 0 \\
0 & - i \end{pmatrix}}_1 {}_{1} \! \left<\phi_{E_1}\right| \otimes \left|\phi_{E_2-E_3}\right>_2 \footnotesize{\begin{pmatrix}
0 & 0 \\
1 & 0 \end{pmatrix}}_2 {}_{2} \! \left<\phi_{E_2}\right| , & \nonumber
\end{flalign}
\begin{flalign}
& \widehat{\rho}_{12}\left[ \widehat{H}_{int}^1,S_{1y}\right] = \frac{4\pi\hbar^2a_{\uparrow\downarrow}}{m} \frac{N}{16} \sum_{\substack{E_a,E_b \\ E_3}} e^{-E_a-E_b} I^{E_a,E_b}_{E_a-E_3,E_b+E_3} \left|\phi_{E_a}\right>_1 \footnotesize{\begin{pmatrix}
i & 1 \\
1 & -i \end{pmatrix}}_1 {}_{1} \! \left<\phi_{E_a-E_3}\right| \otimes \left|\phi_{E_b}\right>_2 \footnotesize{\begin{pmatrix}
i & 0 \\
1 & 0 \end{pmatrix}}_2 {}_{2} \! \left<\phi_{E_b+E_3}\right| . & \nonumber
\end{flalign}
The second term of the interaction Hamiltonian gives:
\begin{flalign}
& \left[ \widehat{H}_{int}^2,S_{1y}\right] = \frac{4\pi\hbar^2a_{\uparrow\downarrow}}{m} \frac{N}{4} \sum_{E_1,E_2,E_3} I^{E_1+E_3,E_2-E_3}_{E_1,E_2}   \left|\phi_{E_1+E_3}\right>_1 \footnotesize{\begin{pmatrix}
i & 0 \\
0 & -i \end{pmatrix}}_1 {}_{1} \! \left<\phi_{E_1}\right| \otimes \left|\phi_{E_2-E_3}\right>_2 \footnotesize{\begin{pmatrix}
0 & 1 \\
0 & 0 \end{pmatrix}}_2 {}_{2} \! \left<\phi_{E_2}\right| , & \nonumber
\end{flalign}
\begin{flalign}
& \widehat{\rho}_{12}\left[ \widehat{H}_{int}^2,S_{1y}\right] = \frac{4\pi\hbar^2a_{\uparrow\downarrow}}{m} \frac{N}{16} \sum_{\substack{E_a,E_b \\ E_3}} e^{-E_a-E_b} I^{E_a,E_b}_{E_a-E_3,E_b+E_3} \left|\phi_{E_a}\right>_1 \footnotesize{\begin{pmatrix}
i & 1 \\
1 & -i \end{pmatrix}}_1 {}_{1} \! \left<\phi_{E_a-E_3}\right| \otimes \left|\phi_{E_b}\right>_2 \footnotesize{\begin{pmatrix}
0 & 1 \\
0 & -i \end{pmatrix}}_2 {}_{2} \! \left<\phi_{E_b+E_3}\right| . & \nonumber
\end{flalign}
The third term of the interaction Hamiltonian gives:
\begin{flalign}
& \left[ \widehat{H}_{int}^3,S_{1y}\right] = \frac{4\pi\hbar^2a_{\uparrow\downarrow}}{m} \frac{N}{4} \sum_{E_1,E_2,E_3} I^{E_1+E_3,E_2-E_3}_{E_1,E_2}  \left|\phi_{E_1+E_3}\right>_1 \footnotesize{\begin{pmatrix}
0 & -i \\
-i & 0 \end{pmatrix}}_1 {}_{1} \! \left<\phi_{E_1}\right| \otimes \left|\phi_{E_2-E_3}\right>_2 \footnotesize{\begin{pmatrix}
0 & 0 \\
0 & 1 \end{pmatrix}}_2 {}_{2} \! \left<\phi_{E_2}\right| ,& \nonumber
\end{flalign}
\begin{flalign}
& \widehat{\rho}_{12}\left[ \widehat{H}_{int}^3,S_{1y}\right] = \frac{4\pi\hbar^2a_{\uparrow\downarrow}}{m} \frac{N}{16} \sum_{\substack{E_a,E_b \\ E_3}} e^{-E_a-E_b} I^{E_a,E_b}_{E_a-E_3,E_b+E_3} \left|\phi_{E_a}\right>_1 \footnotesize{\begin{pmatrix}
1 & -i \\
-i & -1 \end{pmatrix}}_1 {}_{1} \! \left<\phi_{E_a-E_3}\right| \otimes \left|\phi_{E_b}\right>_2 \footnotesize{\begin{pmatrix}
0 & i \\
0 & 1 \end{pmatrix}}_2 {}_{2} \! \left<\phi_{E_b+E_3}\right| . & \nonumber
\end{flalign}
The fourth term of the interaction Hamiltonian gives:
\begin{flalign}
& \left[ \widehat{H}_{int}^4,S_{1y}\right] = \frac{4\pi\hbar^2a_{\uparrow\downarrow}}{m} \frac{N}{4} \sum_{E_1,E_2,E_3} I^{E_1+E_3,E_2-E_3}_{E_1,E_2}   \left|\phi_{E_1+E_3}\right>_1 \footnotesize{\begin{pmatrix}
0 & i \\
i & 0 \end{pmatrix}}_1 {}_{1} \! \left<\phi_{E_1}\right| \otimes \left|\phi_{E_2-E_3}\right>_2 \footnotesize{\begin{pmatrix}
1 & 0 \\
0 & 0 \end{pmatrix}}_2 {}_{2} \! \left<\phi_{E_2}\right| , & \nonumber
\end{flalign}
\begin{flalign}
& \widehat{\rho}_{12}\left[ \widehat{H}_{int}^4,S_{1y}\right] = \frac{4\pi\hbar^2a_{\uparrow\downarrow}}{m} \frac{N}{16} \sum_{\substack{E_a,E_b \\ E_3}} e^{-E_a-E_b} I^{E_a,E_b}_{E_a-E_3,E_b+E_3} \left|\phi_{E_a}\right>_1 \footnotesize{\begin{pmatrix}
-1 & i \\
i & 1 \end{pmatrix}}_1 {}_{1} \! \left<\phi_{E_a-E_3}\right| \otimes \left|\phi_{E_b}\right>_2 \footnotesize{\begin{pmatrix}
1 & 0 \\
-i & 0 \end{pmatrix}}_2 {}_{2} \! \left<\phi_{E_b+E_3}\right| . & \nonumber
\end{flalign}
The fifth term of the interaction Hamiltonian gives:
\begin{flalign}
& \left[ \widehat{H}_{int}^5,S_{1y}\right] = \frac{4\pi\hbar^2a_{\uparrow\uparrow}}{m} \frac{N}{2} \sum_{E_1,E_2,E_3} I^{E_1+E_3,E_2-E_3}_{E_1,E_2} \left|\phi_{E_1+E_3}\right>_1 \footnotesize{\begin{pmatrix}
0 & -i \\
-i & 0 \end{pmatrix}}_1 {}_{1} \! \left<\phi_{E_1}\right| \otimes \left|\phi_{E_2-E_3}\right>_2 \footnotesize{\begin{pmatrix}
1 & 0 \\
0 & 0 \end{pmatrix}}_2 {}_{2} \! \left<\phi_{E_2}\right| , & \nonumber
\end{flalign}
\begin{flalign}
& \widehat{\rho}_{12}\left[ \widehat{H}_{int}^5,S_{1y}\right] = \frac{4\pi\hbar^2a_{\uparrow\uparrow}}{m} \frac{N}{8} \sum_{\substack{E_a,E_b \\ E_3}} e^{-E_a-E_b} I^{E_a,E_b}_{E_a-E_3,E_b+E_3} \left|\phi_{E_a}\right>_1 \footnotesize{\begin{pmatrix}
1 & -i \\
-i & -1 \end{pmatrix}}_1 {}_{1} \! \left<\phi_{E_a-E_3}\right| \otimes \left|\phi_{E_b}\right>_2 \footnotesize{\begin{pmatrix}
1 & 0 \\
-i & 0 \end{pmatrix}}_2 {}_{2} \! \left<\phi_{E_b+E_3}\right| . & \nonumber
\end{flalign}
The sixth term of the interaction Hamiltonian gives:
\begin{flalign}
& \left[ \widehat{H}_{int}^6,S_{1y}\right] = \frac{4\pi\hbar^2a_{\downarrow\downarrow}}{m} \frac{N}{2} \sum_{E_1,E_2,E_3} I^{E_1+E_3,E_2-E_3}_{E_1,E_2}  \left|\phi_{E_1+E_3}\right>_1 \footnotesize{\begin{pmatrix}
0 & i \\
i & 0 \end{pmatrix}}_1 {}_{1} \! \left<\phi_{E_1}\right| \otimes \left|\phi_{E_2-E_3}\right>_2 \footnotesize{\begin{pmatrix}
0 & 0 \\
0 & 1 \end{pmatrix}}_2 {}_{2} \! \left<\phi_{E_2}\right| , & \nonumber
\end{flalign}
\begin{flalign}
& \widehat{\rho}_{12}\left[ \widehat{H}_{int}^6,S_{1y}\right] = \frac{4\pi\hbar^2a_{\downarrow\downarrow}}{m} \frac{N}{8} \sum_{\substack{E_a,E_b \\ E_3}} e^{-E_a-E_b} I^{E_a,E_b}_{E_a-E_3,E_b+E_3} \left|\phi_{E_a}\right>_1 \footnotesize{\begin{pmatrix}
-1 & i \\
i & 1 \end{pmatrix}}_1 {}_{1} \! \left<\phi_{E_a-E_3}\right| \otimes \left|\phi_{E_b}\right>_2 \footnotesize{\begin{pmatrix}
0 & i \\
0 & 1 \end{pmatrix}}_2 {}_{2} \! \left<\phi_{E_b+E_3}\right| . & \nonumber
\end{flalign}

\subsection{Commutator $\left[\widehat{H}_{int},S_{1z}\right]$}
\label{sec_HintSz}

The first term of the interaction Hamiltonian gives:
\begin{flalign}
& \left[ \widehat{H}_{int}^1,S_{1z}\right] = \frac{4\pi\hbar^2a_{\uparrow\downarrow}}{m} \frac{N}{4} \sum_{E_1,E_2,E_3} I^{E_1+E_3,E_2-E_3}_{E_1,E_2}   \left|\phi_{E_1+E_3}\right>_1 \footnotesize{\begin{pmatrix}
0 & -2 \\
0 & 0 \end{pmatrix}}_1 {}_{1} \! \left<\phi_{E_1}\right| \otimes \left|\phi_{E_2-E_3}\right>_2 \footnotesize{\begin{pmatrix}
0 & 0 \\
1 & 0 \end{pmatrix}}_2 {}_{2} \! \left<\phi_{E_2}\right| , & \nonumber
\end{flalign}
\begin{flalign}
& \widehat{\rho}_{12}\left[ \widehat{H}_{int}^1,S_{1z}\right] = \frac{4\pi\hbar^2a_{\uparrow\downarrow}}{m} \frac{N}{16} \sum_{\substack{E_a,E_b \\ E_3}} e^{-E_a-E_b} I^{E_a,E_b}_{E_a-E_3,E_b+E_3} \left|\phi_{E_a}\right>_1 \footnotesize{\begin{pmatrix}
0 & -2 \\
0 & 2i \end{pmatrix}}_1 {}_{1} \! \left<\phi_{E_a-E_3}\right| \otimes \left|\phi_{E_b}\right>_2 \footnotesize{\begin{pmatrix}
i & 0 \\
1 & 0 \end{pmatrix}}_2 {}_{2} \! \left<\phi_{E_b+E_3}\right| . & \nonumber
\end{flalign}
The second term of the interaction Hamiltonian gives:
\begin{flalign}
& \left[ \widehat{H}_{int}^2,S_{1z}\right] = \frac{4\pi\hbar^2a_{\uparrow\downarrow}}{m} \frac{N}{4} \sum_{E_1,E_2,E_3} I^{E_1+E_3,E_2-E_3}_{E_1,E_2}   \left|\phi_{E_1+E_3}\right>_1 \footnotesize{\begin{pmatrix}
0 & 0 \\
2 & 0 \end{pmatrix}}_1 {}_{1} \! \left<\phi_{E_1}\right| \otimes \left|\phi_{E_2-E_3}\right>_2 \footnotesize{\begin{pmatrix}
0 & 1 \\
0 & 0 \end{pmatrix}}_2 {}_{2} \! \left<\phi_{E_2}\right| , & \nonumber
\end{flalign}
\begin{flalign}
& \widehat{\rho}_{12}\left[ \widehat{H}_{int}^2,S_{1z}\right] = \frac{4\pi\hbar^2a_{\uparrow\downarrow}}{m} \frac{N}{16} \sum_{\substack{E_a,E_b \\ E_3}} e^{-E_a-E_b} I^{E_a,E_b}_{E_a-E_3,E_b+E_3} \left|\phi_{E_a}\right>_1 \footnotesize{\begin{pmatrix}
2i & 0 \\
2 & 0 \end{pmatrix}}_1 {}_{1} \! \left<\phi_{E_a-E_3}\right| \otimes \left|\phi_{E_b}\right>_2 \footnotesize{\begin{pmatrix}
0 & 1 \\
0 & -i \end{pmatrix}}_2 {}_{2} \! \left<\phi_{E_b+E_3}\right| . & \nonumber
\end{flalign}
The third term of the interaction Hamiltonian gives:
\begin{flalign}
& \left[ \widehat{H}_{int}^3,S_{1z}\right] = \frac{4\pi\hbar^2a_{\uparrow\downarrow}}{m} \frac{N}{4} \sum_{E_1,E_2,E_3} I^{E_1+E_3,E_2-E_3}_{E_1,E_2}   \left|\phi_{E_1+E_3}\right>_1 \footnotesize{\begin{pmatrix}
0 & 0 \\
0 & 0 \end{pmatrix}}_1 {}_{1} \! \left<\phi_{E_1}\right| \otimes \left|\phi_{E_2-E_3}\right>_2 \footnotesize{\begin{pmatrix}
0 & 0 \\
0 & 1 \end{pmatrix}}_2 {}_{2} \! \left<\phi_{E_2}\right| , & \nonumber
\end{flalign}
\begin{flalign}
& \widehat{\rho}_{12}\left[ \widehat{H}_{int}^3,S_{1z}\right] = 0 . & \nonumber
\end{flalign}
The fourth term of the interaction Hamiltonian gives:
\begin{flalign}
& \left[ \widehat{H}_{int}^4,S_{1z}\right] = \frac{4\pi\hbar^2a_{\uparrow\downarrow}}{m} \frac{N}{4} \sum_{E_1,E_2,E_3} I^{E_1+E_3,E_2-E_3}_{E_1,E_2} \left|\phi_{E_1+E_3}\right>_1 \footnotesize{\begin{pmatrix}
0 & 0 \\
0 & 0 \end{pmatrix}}_1 {}_{1} \! \left<\phi_{E_1}\right| \otimes \left|\phi_{E_2-E_3}\right>_2 \footnotesize{\begin{pmatrix}
1 & 0 \\
0 & 0 \end{pmatrix}}_2 {}_{2} \! \left<\phi_{E_2}\right| , & \nonumber
\end{flalign}
\begin{flalign}
& \widehat{\rho}_{12}\left[ \widehat{H}_{int}^4,S_{1z}\right] = 0 . & \nonumber
\end{flalign}
The fifth term of the interaction Hamiltonian gives:
\begin{flalign}
& \left[ \widehat{H}_{int}^5,S_{1z}\right] = \frac{4\pi\hbar^2a_{\uparrow\uparrow}}{m} \frac{N}{2} \sum_{E_1,E_2,E_3} I^{E_1+E_3,E_2-E_3}_{E_1,E_2}   \left|\phi_{E_1+E_3}\right>_1 \footnotesize{\begin{pmatrix}
0 & 0 \\
0 & 0 \end{pmatrix}}_1 {}_{1} \! \left<\phi_{E_1}\right| \otimes \left|\phi_{E_2-E_3}\right>_2 \footnotesize{\begin{pmatrix}
1 & 0 \\
0 & 0 \end{pmatrix}}_2 {}_{2} \! \left<\phi_{E_2}\right| , & \nonumber
\end{flalign}
\begin{flalign}
& \widehat{\rho}_{12}\left[ \widehat{H}_{int}^5,S_{1z}\right] = 0 . & \nonumber
\end{flalign}
The sixth term of the interaction Hamiltonian gives:
\begin{flalign}
& \left[ \widehat{H}_{int}^6,S_{1z}\right] = \frac{4\pi\hbar^2a_{\downarrow\downarrow}}{m} \frac{N}{2} \sum_{E_1,E_2,E_3} I^{E_1+E_3,E_2-E_3}_{E_1,E_2}  \left|\phi_{E_1+E_3}\right>_1 \footnotesize{\begin{pmatrix}
0 & 0 \\
0 & 0 \end{pmatrix}}_1 {}_{1} \! \left<\phi_{E_1}\right| \otimes \left|\phi_{E_2-E_3}\right>_2 \footnotesize{\begin{pmatrix}
0 & 0 \\
0 & 1 \end{pmatrix}}_2 {}_{2} \! \left<\phi_{E_2}\right| , & \nonumber
\end{flalign}
\begin{flalign}
& \widehat{\rho}_{12}\left[ \widehat{H}_{int}^6,S_{1z}\right] = 0 . & \nonumber
\end{flalign}

We also remark that:
\begin{flalign}
& \begin{pmatrix}
0 & -2 \\
0 & 2i \end{pmatrix}_1 \otimes \begin{pmatrix}
i & 0 \\
1 & 0 \end{pmatrix}_2 + \begin{pmatrix}
2i & 0 \\
2 & 0 \end{pmatrix}_1 \otimes \begin{pmatrix}
0 & 1 \\
0 & -i \end{pmatrix}_2 =  i \left( Id^1 - \sigma_y^1 \right) \otimes \left( \sigma_x^2+i\sigma_z^2 \right) - i \left( \sigma_x^1+i\sigma_z^1 \right) \otimes \left( Id^2-\sigma_y^2 \right) . & \nonumber
\end{flalign}
\end{widetext}

\section{General equation for the dynamics of the one atom average spin}
\label{sec_FullISRE}

In this appendix, we give the equation for the dynamics of the one atom average spin in the case where i) the three interaction lengths $a_{\uparrow\downarrow}$, $a_{\downarrow\downarrow}$ and $a_{\uparrow\uparrow}$ are different and ii) the wave functions are different for the two states, i.e. $\phi_E^\uparrow(r) \neq \phi_E^\downarrow(r)$. To derive this result we follow the same demonstration as in appendix \ref{sec_Demonstration}. The average of the commutator between $\widehat{H}_{int}$ and $\vec{S}_1$ is computed from the results of the appendix \ref{sec_Appendix}. This leads to:
\begin{widetext}
\begin{eqnarray}
\frac{d}{dt} \left<\vec{S}_1(E)\right> & = & \Omega(E) \vec{e}_z \wedge \left<\vec{S}_{1}(E)\right> \nonumber\\
& + &\int dE' \frac{E'^2}{2} f^0_{\uparrow\downarrow}(E,E') \left<\vec{S}_{1}(E')\right> \wedge \left<\vec{S}_{1}(E)\right> \nonumber\\
& + & \int dE' \frac{E'^2}{2} \left[ 2f^+(E,E') - f^+_{\downarrow\uparrow}(E,E') - f^°_{\uparrow\downarrow}(E,E') \right] 
\left<S_{1z}(E')\right> \vec{e}_z \wedge \left<\vec{S}_{1}(E)\right> \nonumber\\
& - & \int dE' \frac{E'^2}{2}g^0_{\uparrow\downarrow}(E,E')  \left\{ \left<S_{1z}(E)\right> \left<\vec{S}_{1}(E')\right> - \left[  \left<\vec{S}_{1}(E)\right> \cdot \left<\vec{S}_{1}(E')\right> \right]\vec{e}_z \right\} \nonumber\\
& - & \int dE' \frac{E'^2}{2} \left[ 2f^-(E,E') + f^-_{\downarrow\uparrow}(E,E') \right] \left[ \left<S_{1x}(E')\right> + \left<S_{1y}(E')\right> \right] \vec{e}_z \wedge \left<\vec{S}_{1}(E)\right>  \nonumber\\
& + &  i\int dE' \frac{E'^2}{2} \left[ 2f^-(E,E')+f^-_{\downarrow\uparrow}(E,E') \right] \left<S_{1z}(E')\right> \vec{e}_z \wedge \left<\vec{S}_{1}(E)\right> \nonumber\\ 
& + & \left< \frac{\partial}{\partial t} \vec{S}_1(E) \right> . 
\end{eqnarray}
We have defined:
\begin{eqnarray}
f^\pm_{\downarrow\uparrow} (E,E') & = & \omega_{\uparrow\downarrow} \frac{f_{\downarrow\uparrow} (E',E) \pm f_{\downarrow\uparrow} (E,E')}{2} \nonumber\\
f^\pm (E,E') & = & \frac{\omega_{\uparrow\uparrow}f_{\uparrow\uparrow} (E,E') \pm \omega_{\downarrow\downarrow}f_{\downarrow\downarrow} (E,E')}{2} \nonumber\\
f^0_{\uparrow\downarrow} (E,E') & = & \omega_{\uparrow\downarrow} f_{\uparrow\downarrow}(E,E') \nonumber\\
g^0_{\uparrow\downarrow} (E,E') & = & \omega_{\uparrow\downarrow} g_{\uparrow\downarrow}(E,E')
\end{eqnarray}
\end{widetext}
where $\omega_{ij} = 4\pi\hbar a_{ij} N /(mV_{eff})$, with $\{i,j\} = \{\uparrow,\downarrow\}$, and:
\begin{eqnarray}
K^{E\uparrow,E'\downarrow}_{E\downarrow,E'\uparrow} & = & f_{\uparrow\downarrow}(E,E') + ig_{\uparrow\downarrow}(E,E') \nonumber\\
K^{E\downarrow,E'\uparrow}_{E\uparrow,E'\downarrow} & = & f_{\uparrow\downarrow}(E,E') - ig_{\uparrow\downarrow}(E,E') \nonumber\\
\end{eqnarray}

\begin{eqnarray}
K^{E\uparrow,E'\downarrow}_{E\uparrow,E'\downarrow} & = & f_{\downarrow\uparrow}(E',E) \qquad K^{E\downarrow,E'\uparrow}_{E\downarrow,E'\uparrow} = f_{\downarrow\uparrow}(E,E') \nonumber\\
K^{E\uparrow,E'\uparrow}_{E\uparrow,E'\uparrow} & = & f_{\uparrow\uparrow}(E,E') \qquad K^{E\downarrow,E'\downarrow}_{E\downarrow,E'\downarrow}  = f_{\downarrow\downarrow}(E,E')
\end{eqnarray}
$f_{\uparrow\downarrow}(E,E')$, $g_{\uparrow\downarrow}(E,E')$, $f_{\downarrow\uparrow}(E,E')$, $f_{\uparrow\uparrow}(E,E')$ and   $f_{\downarrow\downarrow}(E,E')$ are real. We also defined the interaction kernel in a way similar form to equation (\ref{eq_WaveOverLap}):
\begin{eqnarray}
K^{E\uparrow,E'\downarrow}_{E\downarrow,E'\uparrow} & = & \frac{I^{E\uparrow,E'\downarrow}_{E\downarrow,E'\uparrow} }{V_{eff}} \nonumber\\
& = & \int \phi_{E}^{\uparrow*}(r)\phi_{E'}^{\downarrow*}(r)\phi_{E}^\downarrow(r)\phi_{E'}^\uparrow(r) \frac{ dr }{V_{eff}} .
\end{eqnarray}

\bibliography{biblio}

\providecommand{\noopsort}[1]{}\providecommand{\singleletter}[1]{#1}%
\begin{thebibliography}{60}%
\makeatletter
\providecommand \@ifxundefined [1]{%
 \@ifx{#1\undefined}
}%
\providecommand \@ifnum [1]{%
 \ifnum #1\expandafter \@firstoftwo
 \else \expandafter \@secondoftwo
 \fi
}%
\providecommand \@ifx [1]{%
 \ifx #1\expandafter \@firstoftwo
 \else \expandafter \@secondoftwo
 \fi
}%
\providecommand \natexlab [1]{#1}%
\providecommand \enquote  [1]{``#1''}%
\providecommand \bibnamefont  [1]{#1}%
\providecommand \bibfnamefont [1]{#1}%
\providecommand \citenamefont [1]{#1}%
\providecommand \href@noop [0]{\@secondoftwo}%
\providecommand \href [0]{\begingroup \@sanitize@url \@href}%
\providecommand \@href[1]{\@@startlink{#1}\@@href}%
\providecommand \@@href[1]{\endgroup#1\@@endlink}%
\providecommand \@sanitize@url [0]{\catcode `\\12\catcode `\$12\catcode
  `\&12\catcode `\#12\catcode `\^12\catcode `\_12\catcode `\%12\relax}%
\providecommand \@@startlink[1]{}%
\providecommand \@@endlink[0]{}%
\providecommand \url  [0]{\begingroup\@sanitize@url \@url }%
\providecommand \@url [1]{\endgroup\@href {#1}{\urlprefix }}%
\providecommand \urlprefix  [0]{URL }%
\providecommand \Eprint [0]{\href }%
\providecommand \doibase [0]{http://dx.doi.org/}%
\providecommand \selectlanguage [0]{\@gobble}%
\providecommand \bibinfo  [0]{\@secondoftwo}%
\providecommand \bibfield  [0]{\@secondoftwo}%
\providecommand \translation [1]{[#1]}%
\providecommand \BibitemOpen [0]{}%
\providecommand \bibitemStop [0]{}%
\providecommand \bibitemNoStop [0]{.\EOS\space}%
\providecommand \EOS [0]{\spacefactor3000\relax}%
\providecommand \BibitemShut  [1]{\csname bibitem#1\endcsname}%
\let\auto@bib@innerbib\@empty
\bibitem [{\citenamefont {Dupont-Nivet}\ \emph {et~al.}(2018)\citenamefont
  {Dupont-Nivet}, \citenamefont {Demur}, \citenamefont {Westbrook},\ and\
  \citenamefont {Schwartz}}]{DupontNivet2017b}%
  \BibitemOpen
  \bibfield  {author} {\bibinfo {author} {\bibfnamefont {M.}~\bibnamefont
  {Dupont-Nivet}}, \bibinfo {author} {\bibfnamefont {R.}~\bibnamefont {Demur}},
  \bibinfo {author} {\bibfnamefont {C.~I.}\ \bibnamefont {Westbrook}}, \ and\
  \bibinfo {author} {\bibfnamefont {S.}~\bibnamefont {Schwartz}},\ }\href@noop
  {} {\bibfield  {journal} {\bibinfo  {journal} {New J. Phys.}\ }\textbf
  {\bibinfo {volume} {20}},\ \bibinfo {pages} {043051} (\bibinfo {year}
  {2018})}\BibitemShut {NoStop}%
\bibitem [{\citenamefont {Treutlein}\ \emph {et~al.}(2004)\citenamefont
  {Treutlein}, \citenamefont {Hommelhoff}, \citenamefont {Steinmetz},
  \citenamefont {H\"ansch},\ and\ \citenamefont {Reichel}}]{Treutlein2004}%
  \BibitemOpen
  \bibfield  {author} {\bibinfo {author} {\bibfnamefont {P.}~\bibnamefont
  {Treutlein}}, \bibinfo {author} {\bibfnamefont {P.}~\bibnamefont
  {Hommelhoff}}, \bibinfo {author} {\bibfnamefont {T.}~\bibnamefont
  {Steinmetz}}, \bibinfo {author} {\bibfnamefont {T.~W.}\ \bibnamefont
  {H\"ansch}}, \ and\ \bibinfo {author} {\bibfnamefont {J.}~\bibnamefont
  {Reichel}},\ }\href {\doibase 10.1103/PhysRevLett.92.203005} {\bibfield
  {journal} {\bibinfo  {journal} {Phys. Rev. Lett.}\ }\textbf {\bibinfo
  {volume} {92}},\ \bibinfo {pages} {203005} (\bibinfo {year}
  {2004})}\BibitemShut {NoStop}%
\bibitem [{\citenamefont {Deutsch}\ \emph {et~al.}(2010)\citenamefont
  {Deutsch}, \citenamefont {Ramirez-Martinez}, \citenamefont {Lacro\^ute},
  \citenamefont {Reinhard}, \citenamefont {Schneider}, \citenamefont {Fuchs},
  \citenamefont {Pi\'echon}, \citenamefont {Lalo\"e}, \citenamefont {Reichel},\
  and\ \citenamefont {Rosenbusch}}]{Deutsch2010}%
  \BibitemOpen
  \bibfield  {author} {\bibinfo {author} {\bibfnamefont {C.}~\bibnamefont
  {Deutsch}}, \bibinfo {author} {\bibfnamefont {F.}~\bibnamefont
  {Ramirez-Martinez}}, \bibinfo {author} {\bibfnamefont {C.}~\bibnamefont
  {Lacro\^ute}}, \bibinfo {author} {\bibfnamefont {F.}~\bibnamefont
  {Reinhard}}, \bibinfo {author} {\bibfnamefont {T.}~\bibnamefont {Schneider}},
  \bibinfo {author} {\bibfnamefont {J.~N.}\ \bibnamefont {Fuchs}}, \bibinfo
  {author} {\bibfnamefont {F.}~\bibnamefont {Pi\'echon}}, \bibinfo {author}
  {\bibfnamefont {F.}~\bibnamefont {Lalo\"e}}, \bibinfo {author} {\bibfnamefont
  {J.}~\bibnamefont {Reichel}}, \ and\ \bibinfo {author} {\bibfnamefont
  {P.}~\bibnamefont {Rosenbusch}},\ }\href {\doibase
  10.1103/PhysRevLett.105.020401} {\bibfield  {journal} {\bibinfo  {journal}
  {Phys. Rev. Lett.}\ }\textbf {\bibinfo {volume} {105}},\ \bibinfo {pages}
  {020401} (\bibinfo {year} {2010})}\BibitemShut {NoStop}%
\bibitem [{\citenamefont {Szmuk}\ \emph {et~al.}(2015)\citenamefont {Szmuk},
  \citenamefont {Dugrain}, \citenamefont {Maineult}, \citenamefont {Reichel},\
  and\ \citenamefont {Rosenbusch}}]{Szmuk2015}%
  \BibitemOpen
  \bibfield  {author} {\bibinfo {author} {\bibfnamefont {R.}~\bibnamefont
  {Szmuk}}, \bibinfo {author} {\bibfnamefont {V.}~\bibnamefont {Dugrain}},
  \bibinfo {author} {\bibfnamefont {W.}~\bibnamefont {Maineult}}, \bibinfo
  {author} {\bibfnamefont {J.}~\bibnamefont {Reichel}}, \ and\ \bibinfo
  {author} {\bibfnamefont {P.}~\bibnamefont {Rosenbusch}},\ }\href@noop {}
  {\bibfield  {journal} {\bibinfo  {journal} {Phys. Rev. A}\ }\textbf {\bibinfo
  {volume} {92}},\ \bibinfo {pages} {012106} (\bibinfo {year}
  {2015})}\BibitemShut {NoStop}%
\bibitem [{\citenamefont {Ammar}\ \emph {et~al.}(2015)\citenamefont {Ammar},
  \citenamefont {Dupont-Nivet}, \citenamefont {Huet}, \citenamefont {Pocholle},
  \citenamefont {Rosenbusch}, \citenamefont {Bouchoule}, \citenamefont
  {Westbrook}, \citenamefont {Est\`eve}, \citenamefont {Reichel}, \citenamefont
  {Guerlin},\ and\ \citenamefont {Schwartz}}]{Ammar2014}%
  \BibitemOpen
  \bibfield  {author} {\bibinfo {author} {\bibfnamefont {M.}~\bibnamefont
  {Ammar}}, \bibinfo {author} {\bibfnamefont {M.}~\bibnamefont {Dupont-Nivet}},
  \bibinfo {author} {\bibfnamefont {L.}~\bibnamefont {Huet}}, \bibinfo {author}
  {\bibfnamefont {J.-P.}\ \bibnamefont {Pocholle}}, \bibinfo {author}
  {\bibfnamefont {P.}~\bibnamefont {Rosenbusch}}, \bibinfo {author}
  {\bibfnamefont {I.}~\bibnamefont {Bouchoule}}, \bibinfo {author}
  {\bibfnamefont {C.~I.}\ \bibnamefont {Westbrook}}, \bibinfo {author}
  {\bibfnamefont {J.}~\bibnamefont {Est\`eve}}, \bibinfo {author}
  {\bibfnamefont {J.}~\bibnamefont {Reichel}}, \bibinfo {author} {\bibfnamefont
  {C.}~\bibnamefont {Guerlin}}, \ and\ \bibinfo {author} {\bibfnamefont
  {S.}~\bibnamefont {Schwartz}},\ }\href {\doibase 10.1103/PhysRevA.91.053623}
  {\bibfield  {journal} {\bibinfo  {journal} {Phys. Rev. A}\ }\textbf {\bibinfo
  {volume} {91}},\ \bibinfo {pages} {053623} (\bibinfo {year}
  {2015})}\BibitemShut {NoStop}%
\bibitem [{\citenamefont {Dupont-Nivet}\ \emph {et~al.}(2016)\citenamefont
  {Dupont-Nivet}, \citenamefont {Westbrook},\ and\ \citenamefont
  {Schwartz}}]{DupontNivet2014}%
  \BibitemOpen
  \bibfield  {author} {\bibinfo {author} {\bibfnamefont {M.}~\bibnamefont
  {Dupont-Nivet}}, \bibinfo {author} {\bibfnamefont {C.~I.}\ \bibnamefont
  {Westbrook}}, \ and\ \bibinfo {author} {\bibfnamefont {S.}~\bibnamefont
  {Schwartz}},\ }\href@noop {} {\bibfield  {journal} {\bibinfo  {journal} {New
  J. Phys.}\ }\textbf {\bibinfo {volume} {18}},\ \bibinfo {pages} {113012}
  (\bibinfo {year} {2016})}\BibitemShut {NoStop}%
\bibitem [{\citenamefont {Pelle}\ \emph {et~al.}(2013)\citenamefont {Pelle},
  \citenamefont {Hilico}, \citenamefont {Tackmann}, \citenamefont {Beaufils},\
  and\ \citenamefont {Pereira~dos Santos}}]{Pelle2013}%
  \BibitemOpen
  \bibfield  {author} {\bibinfo {author} {\bibfnamefont {B.}~\bibnamefont
  {Pelle}}, \bibinfo {author} {\bibfnamefont {A.}~\bibnamefont {Hilico}},
  \bibinfo {author} {\bibfnamefont {G.}~\bibnamefont {Tackmann}}, \bibinfo
  {author} {\bibfnamefont {Q.}~\bibnamefont {Beaufils}}, \ and\ \bibinfo
  {author} {\bibfnamefont {F.}~\bibnamefont {Pereira~dos Santos}},\ }\href
  {\doibase 10.1103/PhysRevA.87.023601} {\bibfield  {journal} {\bibinfo
  {journal} {Phys. Rev. A}\ }\textbf {\bibinfo {volume} {87}},\ \bibinfo
  {pages} {023601} (\bibinfo {year} {2013})}\BibitemShut {NoStop}%
\bibitem [{\citenamefont {Alauze}\ \emph {et~al.}(2018)\citenamefont {Alauze},
  \citenamefont {Bonnin}, \citenamefont {Solaro},\ and\ \citenamefont
  {Dos~Santos}}]{Alauze2018}%
  \BibitemOpen
  \bibfield  {author} {\bibinfo {author} {\bibfnamefont {X.}~\bibnamefont
  {Alauze}}, \bibinfo {author} {\bibfnamefont {A.}~\bibnamefont {Bonnin}},
  \bibinfo {author} {\bibfnamefont {C.}~\bibnamefont {Solaro}}, \ and\ \bibinfo
  {author} {\bibfnamefont {F.~P.}\ \bibnamefont {Dos~Santos}},\ }\href@noop {}
  {\bibfield  {journal} {\bibinfo  {journal} {New Journal of Physics}\ }\textbf
  {\bibinfo {volume} {20}},\ \bibinfo {pages} {083014} (\bibinfo {year}
  {2018})}\BibitemShut {NoStop}%
\bibitem [{\citenamefont {Xu}\ \emph {et~al.}(2019)\citenamefont {Xu},
  \citenamefont {Jaffe}, \citenamefont {Panda}, \citenamefont {Kristensen},
  \citenamefont {Clark},\ and\ \citenamefont {M{\"u}ller}}]{Xu2019}%
  \BibitemOpen
  \bibfield  {author} {\bibinfo {author} {\bibfnamefont {V.}~\bibnamefont
  {Xu}}, \bibinfo {author} {\bibfnamefont {M.}~\bibnamefont {Jaffe}}, \bibinfo
  {author} {\bibfnamefont {C.~D.}\ \bibnamefont {Panda}}, \bibinfo {author}
  {\bibfnamefont {S.~L.}\ \bibnamefont {Kristensen}}, \bibinfo {author}
  {\bibfnamefont {L.~W.}\ \bibnamefont {Clark}}, \ and\ \bibinfo {author}
  {\bibfnamefont {H.}~\bibnamefont {M{\"u}ller}},\ }\href@noop {} {\bibfield
  {journal} {\bibinfo  {journal} {Science}\ }\textbf {\bibinfo {volume}
  {366}},\ \bibinfo {pages} {745} (\bibinfo {year} {2019})}\BibitemShut
  {NoStop}%
\bibitem [{\citenamefont {Alzar}\ \emph {et~al.}(2012)\citenamefont {Alzar},
  \citenamefont {Yan},\ and\ \citenamefont {Landragin}}]{GarridoAlzar2012}%
  \BibitemOpen
  \bibfield  {author} {\bibinfo {author} {\bibfnamefont {C.~L.~G.}\
  \bibnamefont {Alzar}}, \bibinfo {author} {\bibfnamefont {W.}~\bibnamefont
  {Yan}}, \ and\ \bibinfo {author} {\bibfnamefont {A.}~\bibnamefont
  {Landragin}},\ }in\ \href {\doibase 10.1364/HILAS.2012.JT2A.10} {\emph
  {\bibinfo {booktitle} {Research in Optical Sciences}}}\ (\bibinfo
  {publisher} {Optical Society of America},\ \bibinfo {year} {2012})\ p.\
  \bibinfo {pages} {JT2A.10}\BibitemShut {NoStop}%
\bibitem [{\citenamefont {Moan}\ \emph {et~al.}(2019)\citenamefont {Moan},
  \citenamefont {Horne}, \citenamefont {Arpornthip}, \citenamefont {Luo},
  \citenamefont {Fallon}, \citenamefont {Berl},\ and\ \citenamefont
  {Sackett}}]{Moan2019}%
  \BibitemOpen
  \bibfield  {author} {\bibinfo {author} {\bibfnamefont {E.}~\bibnamefont
  {Moan}}, \bibinfo {author} {\bibfnamefont {R.}~\bibnamefont {Horne}},
  \bibinfo {author} {\bibfnamefont {T.}~\bibnamefont {Arpornthip}}, \bibinfo
  {author} {\bibfnamefont {Z.}~\bibnamefont {Luo}}, \bibinfo {author}
  {\bibfnamefont {A.}~\bibnamefont {Fallon}}, \bibinfo {author} {\bibfnamefont
  {S.}~\bibnamefont {Berl}}, \ and\ \bibinfo {author} {\bibfnamefont
  {C.}~\bibnamefont {Sackett}},\ }\href@noop {} {\bibfield  {journal} {\bibinfo
   {journal} {arXiv preprint arXiv:1907.05466}\ } (\bibinfo {year}
  {2019})}\BibitemShut {NoStop}%
\bibitem [{\citenamefont {Sadgrove}\ \emph {et~al.}(2013)\citenamefont
  {Sadgrove}, \citenamefont {Eto}, \citenamefont {Sekine}, \citenamefont
  {Suzuki},\ and\ \citenamefont {Hirano}}]{Sadgrove2013}%
  \BibitemOpen
  \bibfield  {author} {\bibinfo {author} {\bibfnamefont {M.}~\bibnamefont
  {Sadgrove}}, \bibinfo {author} {\bibfnamefont {Y.}~\bibnamefont {Eto}},
  \bibinfo {author} {\bibfnamefont {S.}~\bibnamefont {Sekine}}, \bibinfo
  {author} {\bibfnamefont {H.}~\bibnamefont {Suzuki}}, \ and\ \bibinfo {author}
  {\bibfnamefont {T.}~\bibnamefont {Hirano}},\ }\href@noop {} {\bibfield
  {journal} {\bibinfo  {journal} {J. Phys. Soc. Jpn.}\ }\textbf {\bibinfo
  {volume} {82}},\ \bibinfo {pages} {094002} (\bibinfo {year}
  {2013})}\BibitemShut {NoStop}%
\bibitem [{\citenamefont {Eto}\ \emph {et~al.}(2016)\citenamefont {Eto},
  \citenamefont {Sadrove},\ and\ \citenamefont {Hirano}}]{Eto2016}%
  \BibitemOpen
  \bibfield  {author} {\bibinfo {author} {\bibfnamefont {Y.}~\bibnamefont
  {Eto}}, \bibinfo {author} {\bibfnamefont {M.}~\bibnamefont {Sadrove}}, \ and\
  \bibinfo {author} {\bibfnamefont {T.}~\bibnamefont {Hirano}},\ }in\
  \href@noop {} {\emph {\bibinfo {booktitle} {Principles and Methods of Quantum
  Information Technologies}}}\ (\bibinfo  {publisher} {Springer},\ \bibinfo
  {year} {2016})\ pp.\ \bibinfo {pages} {111--133}\BibitemShut {NoStop}%
\bibitem [{\citenamefont {Grond}\ \emph {et~al.}(2010)\citenamefont {Grond},
  \citenamefont {Hohenester}, \citenamefont {Mazets},\ and\ \citenamefont
  {Schmiedmayer}}]{Grond2010}%
  \BibitemOpen
  \bibfield  {author} {\bibinfo {author} {\bibfnamefont {J.}~\bibnamefont
  {Grond}}, \bibinfo {author} {\bibfnamefont {U.}~\bibnamefont {Hohenester}},
  \bibinfo {author} {\bibfnamefont {I.}~\bibnamefont {Mazets}}, \ and\ \bibinfo
  {author} {\bibfnamefont {J.}~\bibnamefont {Schmiedmayer}},\ }\href
  {http://stacks.iop.org/1367-2630/12/i=6/a=065036} {\bibfield  {journal}
  {\bibinfo  {journal} {New J. Phys.}\ }\textbf {\bibinfo {volume} {12}},\
  \bibinfo {pages} {065036} (\bibinfo {year} {2010})}\BibitemShut {NoStop}%
\bibitem [{\citenamefont {Abend}\ \emph {et~al.}(2016)\citenamefont {Abend},
  \citenamefont {Gebbe}, \citenamefont {Gersemann}, \citenamefont {Ahlers},
  \citenamefont {M{\"u}ntinga}, \citenamefont {Giese}, \citenamefont {Gaaloul},
  \citenamefont {Schubert}, \citenamefont {L{\"a}mmerzahl}, \citenamefont
  {Ertmer}, \citenamefont {Schleich},\ and\ \citenamefont {Rasel}}]{Abend2016}%
  \BibitemOpen
  \bibfield  {author} {\bibinfo {author} {\bibfnamefont {S.}~\bibnamefont
  {Abend}}, \bibinfo {author} {\bibfnamefont {M.}~\bibnamefont {Gebbe}},
  \bibinfo {author} {\bibfnamefont {M.}~\bibnamefont {Gersemann}}, \bibinfo
  {author} {\bibfnamefont {H.}~\bibnamefont {Ahlers}}, \bibinfo {author}
  {\bibfnamefont {H.}~\bibnamefont {M{\"u}ntinga}}, \bibinfo {author}
  {\bibfnamefont {E.}~\bibnamefont {Giese}}, \bibinfo {author} {\bibfnamefont
  {N.}~\bibnamefont {Gaaloul}}, \bibinfo {author} {\bibfnamefont
  {C.}~\bibnamefont {Schubert}}, \bibinfo {author} {\bibfnamefont
  {C.}~\bibnamefont {L{\"a}mmerzahl}}, \bibinfo {author} {\bibfnamefont
  {W.}~\bibnamefont {Ertmer}}, \bibinfo {author} {\bibfnamefont {W.~P.}\
  \bibnamefont {Schleich}}, \ and\ \bibinfo {author} {\bibfnamefont {E.~M.}\
  \bibnamefont {Rasel}},\ }\href@noop {} {\bibfield  {journal} {\bibinfo
  {journal} {Phys. Rev. Lett.}\ }\textbf {\bibinfo {volume} {117}},\ \bibinfo
  {pages} {203003} (\bibinfo {year} {2016})}\BibitemShut {NoStop}%
\bibitem [{\citenamefont {Schumm}\ \emph {et~al.}(2005)\citenamefont {Schumm},
  \citenamefont {Hofferberth}, \citenamefont {Andersson}, \citenamefont
  {Wildermuth}, \citenamefont {Groth}, \citenamefont {Bar-Joseph},
  \citenamefont {Schmiedmayer},\ and\ \citenamefont {Kruger}}]{Schumm2005}%
  \BibitemOpen
  \bibfield  {author} {\bibinfo {author} {\bibfnamefont {T.}~\bibnamefont
  {Schumm}}, \bibinfo {author} {\bibfnamefont {S.}~\bibnamefont {Hofferberth}},
  \bibinfo {author} {\bibfnamefont {L.~M.}\ \bibnamefont {Andersson}}, \bibinfo
  {author} {\bibfnamefont {S.}~\bibnamefont {Wildermuth}}, \bibinfo {author}
  {\bibfnamefont {S.}~\bibnamefont {Groth}}, \bibinfo {author} {\bibfnamefont
  {I.}~\bibnamefont {Bar-Joseph}}, \bibinfo {author} {\bibfnamefont
  {J.}~\bibnamefont {Schmiedmayer}}, \ and\ \bibinfo {author} {\bibfnamefont
  {P.}~\bibnamefont {Kruger}},\ }\href@noop {} {\bibfield  {journal} {\bibinfo
  {journal} {Nat. Phys.}\ }\textbf {\bibinfo {volume} {1}},\ \bibinfo {pages}
  {57} (\bibinfo {year} {2005})}\BibitemShut {NoStop}%
\bibitem [{\citenamefont {Javanainen}\ and\ \citenamefont
  {Wilkens}(1997)}]{Javanainen1997}%
  \BibitemOpen
  \bibfield  {author} {\bibinfo {author} {\bibfnamefont {J.}~\bibnamefont
  {Javanainen}}\ and\ \bibinfo {author} {\bibfnamefont {M.}~\bibnamefont
  {Wilkens}},\ }\href {\doibase 10.1103/PhysRevLett.78.4675} {\bibfield
  {journal} {\bibinfo  {journal} {Phys. Rev. Lett.}\ }\textbf {\bibinfo
  {volume} {78}},\ \bibinfo {pages} {4675} (\bibinfo {year}
  {1997})}\BibitemShut {NoStop}%
\bibitem [{\citenamefont {Jo}\ \emph {et~al.}(2007)\citenamefont {Jo},
  \citenamefont {Shin}, \citenamefont {Will}, \citenamefont {Pasquini},
  \citenamefont {Saba}, \citenamefont {Ketterle}, \citenamefont {Pritchard},
  \citenamefont {Vengalattore},\ and\ \citenamefont {Prentiss}}]{Jo2007}%
  \BibitemOpen
  \bibfield  {author} {\bibinfo {author} {\bibfnamefont {G.-B.}\ \bibnamefont
  {Jo}}, \bibinfo {author} {\bibfnamefont {Y.}~\bibnamefont {Shin}}, \bibinfo
  {author} {\bibfnamefont {S.}~\bibnamefont {Will}}, \bibinfo {author}
  {\bibfnamefont {T.~A.}\ \bibnamefont {Pasquini}}, \bibinfo {author}
  {\bibfnamefont {M.}~\bibnamefont {Saba}}, \bibinfo {author} {\bibfnamefont
  {W.}~\bibnamefont {Ketterle}}, \bibinfo {author} {\bibfnamefont {D.~E.}\
  \bibnamefont {Pritchard}}, \bibinfo {author} {\bibfnamefont {M.}~\bibnamefont
  {Vengalattore}}, \ and\ \bibinfo {author} {\bibfnamefont {M.}~\bibnamefont
  {Prentiss}},\ }\href {\doibase 10.1103/PhysRevLett.98.030407} {\bibfield
  {journal} {\bibinfo  {journal} {Phys. Rev. Lett.}\ }\textbf {\bibinfo
  {volume} {98}},\ \bibinfo {pages} {030407} (\bibinfo {year}
  {2007})}\BibitemShut {NoStop}%
\bibitem [{\citenamefont {Berrada}\ \emph {et~al.}(2013)\citenamefont
  {Berrada}, \citenamefont {van Frank}, \citenamefont {B{\"u}cker},
  \citenamefont {Schumm}, \citenamefont {Schaff},\ and\ \citenamefont
  {Schmiedmayer}}]{Berrada2013}%
  \BibitemOpen
  \bibfield  {author} {\bibinfo {author} {\bibfnamefont {T.}~\bibnamefont
  {Berrada}}, \bibinfo {author} {\bibfnamefont {S.}~\bibnamefont {van Frank}},
  \bibinfo {author} {\bibfnamefont {R.}~\bibnamefont {B{\"u}cker}}, \bibinfo
  {author} {\bibfnamefont {T.}~\bibnamefont {Schumm}}, \bibinfo {author}
  {\bibfnamefont {J.-F.}\ \bibnamefont {Schaff}}, \ and\ \bibinfo {author}
  {\bibfnamefont {J.}~\bibnamefont {Schmiedmayer}},\ }\href@noop {} {\bibfield
  {journal} {\bibinfo  {journal} {Nat. Commun.}\ }\textbf {\bibinfo {volume}
  {4}} (\bibinfo {year} {2013})}\BibitemShut {NoStop}%
\bibitem [{\citenamefont {Hosten}\ \emph {et~al.}(2016)\citenamefont {Hosten},
  \citenamefont {Engelsen}, \citenamefont {Krishnakumar},\ and\ \citenamefont
  {Kasevich}}]{Hosten2016}%
  \BibitemOpen
  \bibfield  {author} {\bibinfo {author} {\bibfnamefont {O.}~\bibnamefont
  {Hosten}}, \bibinfo {author} {\bibfnamefont {N.}~\bibnamefont {Engelsen}},
  \bibinfo {author} {\bibfnamefont {R.}~\bibnamefont {Krishnakumar}}, \ and\
  \bibinfo {author} {\bibfnamefont {M.}~\bibnamefont {Kasevich}},\ }\href@noop
  {} {\bibfield  {journal} {\bibinfo  {journal} {Nature}\ } (\bibinfo {year}
  {2016})}\BibitemShut {NoStop}%
\bibitem [{\citenamefont {Haas}\ \emph {et~al.}(2014)\citenamefont {Haas},
  \citenamefont {Volz}, \citenamefont {Gehr}, \citenamefont {Reichel},\ and\
  \citenamefont {Est{\`e}ve}}]{Haas2014b}%
  \BibitemOpen
  \bibfield  {author} {\bibinfo {author} {\bibfnamefont {F.}~\bibnamefont
  {Haas}}, \bibinfo {author} {\bibfnamefont {J.}~\bibnamefont {Volz}}, \bibinfo
  {author} {\bibfnamefont {R.}~\bibnamefont {Gehr}}, \bibinfo {author}
  {\bibfnamefont {J.}~\bibnamefont {Reichel}}, \ and\ \bibinfo {author}
  {\bibfnamefont {J.}~\bibnamefont {Est{\`e}ve}},\ }\href@noop {} {\bibfield
  {journal} {\bibinfo  {journal} {Science}\ }\textbf {\bibinfo {volume}
  {344}},\ \bibinfo {pages} {180} (\bibinfo {year} {2014})}\BibitemShut
  {NoStop}%
\bibitem [{\citenamefont {Barontini}\ \emph {et~al.}(2015)\citenamefont
  {Barontini}, \citenamefont {Hohmann}, \citenamefont {Haas}, \citenamefont
  {Est{\`e}ve},\ and\ \citenamefont {Reichel}}]{Barontini2015}%
  \BibitemOpen
  \bibfield  {author} {\bibinfo {author} {\bibfnamefont {G.}~\bibnamefont
  {Barontini}}, \bibinfo {author} {\bibfnamefont {L.}~\bibnamefont {Hohmann}},
  \bibinfo {author} {\bibfnamefont {F.}~\bibnamefont {Haas}}, \bibinfo {author}
  {\bibfnamefont {J.}~\bibnamefont {Est{\`e}ve}}, \ and\ \bibinfo {author}
  {\bibfnamefont {J.}~\bibnamefont {Reichel}},\ }\href@noop {} {\bibfield
  {journal} {\bibinfo  {journal} {Science}\ }\textbf {\bibinfo {volume}
  {349}},\ \bibinfo {pages} {1317} (\bibinfo {year} {2015})}\BibitemShut
  {NoStop}%
\bibitem [{\citenamefont {Kleine~B\"uning}\ \emph {et~al.}(2011)\citenamefont
  {Kleine~B\"uning}, \citenamefont {Will}, \citenamefont {Ertmer},
  \citenamefont {Rasel}, \citenamefont {Arlt}, \citenamefont {Klempt},
  \citenamefont {Ramirez-Martinez}, \citenamefont {Pi\'echon},\ and\
  \citenamefont {Rosenbusch}}]{Kleine2011}%
  \BibitemOpen
  \bibfield  {author} {\bibinfo {author} {\bibfnamefont {G.}~\bibnamefont
  {Kleine~B\"uning}}, \bibinfo {author} {\bibfnamefont {J.}~\bibnamefont
  {Will}}, \bibinfo {author} {\bibfnamefont {W.}~\bibnamefont {Ertmer}},
  \bibinfo {author} {\bibfnamefont {E.}~\bibnamefont {Rasel}}, \bibinfo
  {author} {\bibfnamefont {J.}~\bibnamefont {Arlt}}, \bibinfo {author}
  {\bibfnamefont {C.}~\bibnamefont {Klempt}}, \bibinfo {author} {\bibfnamefont
  {F.}~\bibnamefont {Ramirez-Martinez}}, \bibinfo {author} {\bibfnamefont
  {F.}~\bibnamefont {Pi\'echon}}, \ and\ \bibinfo {author} {\bibfnamefont
  {P.}~\bibnamefont {Rosenbusch}},\ }\href {\doibase
  10.1103/PhysRevLett.106.240801} {\bibfield  {journal} {\bibinfo  {journal}
  {Phys. Rev. Lett.}\ }\textbf {\bibinfo {volume} {106}},\ \bibinfo {pages}
  {240801} (\bibinfo {year} {2011})}\BibitemShut {NoStop}%
\bibitem [{\citenamefont {Solaro}\ \emph {et~al.}(2016)\citenamefont {Solaro},
  \citenamefont {Bonnin}, \citenamefont {Combes}, \citenamefont {Lopez},
  \citenamefont {Alauze}, \citenamefont {Fuchs}, \citenamefont {Pi{\'e}chon},\
  and\ \citenamefont {Dos~Santos}}]{Solaro2016}%
  \BibitemOpen
  \bibfield  {author} {\bibinfo {author} {\bibfnamefont {C.}~\bibnamefont
  {Solaro}}, \bibinfo {author} {\bibfnamefont {A.}~\bibnamefont {Bonnin}},
  \bibinfo {author} {\bibfnamefont {F.}~\bibnamefont {Combes}}, \bibinfo
  {author} {\bibfnamefont {M.}~\bibnamefont {Lopez}}, \bibinfo {author}
  {\bibfnamefont {X.}~\bibnamefont {Alauze}}, \bibinfo {author} {\bibfnamefont
  {J.-N.}\ \bibnamefont {Fuchs}}, \bibinfo {author} {\bibfnamefont
  {F.}~\bibnamefont {Pi{\'e}chon}}, \ and\ \bibinfo {author} {\bibfnamefont
  {F.~P.}\ \bibnamefont {Dos~Santos}},\ }\href@noop {} {\bibfield  {journal}
  {\bibinfo  {journal} {Phys. Rev. Lett.}\ }\textbf {\bibinfo {volume} {117}},\
  \bibinfo {pages} {163003} (\bibinfo {year} {2016})}\BibitemShut {NoStop}%
\bibitem [{\citenamefont {Deutsch}\ \emph {et~al.}(2012)\citenamefont
  {Deutsch}, \citenamefont {Ramirez-Martinez}, \citenamefont {Lacro\^ute},
  \citenamefont {Maineult}, \citenamefont {Reinhard}, \citenamefont
  {Schneider}, \citenamefont {Fuchs}, \citenamefont {Pi\'echon}, \citenamefont
  {Lalo\"e}, \citenamefont {Reichel},\ and\ \citenamefont
  {Rosenbusch}}]{Deutsch2012}%
  \BibitemOpen
  \bibfield  {author} {\bibinfo {author} {\bibfnamefont {C.}~\bibnamefont
  {Deutsch}}, \bibinfo {author} {\bibfnamefont {F.}~\bibnamefont
  {Ramirez-Martinez}}, \bibinfo {author} {\bibfnamefont {C.}~\bibnamefont
  {Lacro\^ute}}, \bibinfo {author} {\bibfnamefont {W.}~\bibnamefont
  {Maineult}}, \bibinfo {author} {\bibfnamefont {F.}~\bibnamefont {Reinhard}},
  \bibinfo {author} {\bibfnamefont {T.}~\bibnamefont {Schneider}}, \bibinfo
  {author} {\bibfnamefont {J.-N.}\ \bibnamefont {Fuchs}}, \bibinfo {author}
  {\bibfnamefont {F.}~\bibnamefont {Pi\'echon}}, \bibinfo {author}
  {\bibfnamefont {F.}~\bibnamefont {Lalo\"e}}, \bibinfo {author} {\bibfnamefont
  {J.}~\bibnamefont {Reichel}}, \ and\ \bibinfo {author} {\bibfnamefont
  {P.}~\bibnamefont {Rosenbusch}},\ }\href@noop {} {\bibfield  {journal}
  {\bibinfo  {journal} {Revue fran{\c{c}}aise de m{\'e}trologie}\ }\textbf
  {\bibinfo {volume} {2012--1}},\ \bibinfo {pages} {3} (\bibinfo {year}
  {2012})}\BibitemShut {NoStop}%
\bibitem [{\citenamefont {Dupont-Nivet}(2016)}]{DupontNivet2016}%
  \BibitemOpen
  \bibfield  {author} {\bibinfo {author} {\bibfnamefont {M.}~\bibnamefont
  {Dupont-Nivet}},\ }\emph {\bibinfo {title} {Vers un acc\'el\'erom\'etre
  atomique sur puce}},\ \href@noop {} {Ph.D. thesis},\ \bibinfo  {school}
  {Universit\'e Paris Saclay} (\bibinfo {year} {2016})\BibitemShut {NoStop}%
\bibitem [{\citenamefont {B{\"o}hi}\ \emph {et~al.}(2009)\citenamefont
  {B{\"o}hi}, \citenamefont {Riedel}, \citenamefont {Hoffrogge}, \citenamefont
  {Reichel}, \citenamefont {Hansch},\ and\ \citenamefont
  {Treutlein}}]{Bohi2009}%
  \BibitemOpen
  \bibfield  {author} {\bibinfo {author} {\bibfnamefont {P.}~\bibnamefont
  {B{\"o}hi}}, \bibinfo {author} {\bibfnamefont {M.}~\bibnamefont {Riedel}},
  \bibinfo {author} {\bibfnamefont {J.}~\bibnamefont {Hoffrogge}}, \bibinfo
  {author} {\bibfnamefont {J.}~\bibnamefont {Reichel}}, \bibinfo {author}
  {\bibfnamefont {T.}~\bibnamefont {Hansch}}, \ and\ \bibinfo {author}
  {\bibfnamefont {P.}~\bibnamefont {Treutlein}},\ }\href@noop {} {\bibfield
  {journal} {\bibinfo  {journal} {Nat. Phys.}\ }\textbf {\bibinfo {volume}
  {5}},\ \bibinfo {pages} {592} (\bibinfo {year} {2009})}\BibitemShut {NoStop}%
\bibitem [{\citenamefont {Lalo{\"e}}\ and\ \citenamefont
  {Freed}(1988)}]{Laloe1988}%
  \BibitemOpen
  \bibfield  {author} {\bibinfo {author} {\bibfnamefont {F.}~\bibnamefont
  {Lalo{\"e}}}\ and\ \bibinfo {author} {\bibfnamefont {J.~H.}\ \bibnamefont
  {Freed}},\ }\href@noop {} {\bibfield  {journal} {\bibinfo  {journal} {Sci.
  Am.}\ }\textbf {\bibinfo {volume} {258}},\ \bibinfo {pages} {94} (\bibinfo
  {year} {1988})}\BibitemShut {NoStop}%
\bibitem [{\citenamefont {Lhuillier}\ and\ \citenamefont
  {Lalo{\"e}}(1982{\natexlab{a}})}]{Lhuillier1982b}%
  \BibitemOpen
  \bibfield  {author} {\bibinfo {author} {\bibfnamefont {C.}~\bibnamefont
  {Lhuillier}}\ and\ \bibinfo {author} {\bibfnamefont {F.}~\bibnamefont
  {Lalo{\"e}}},\ }\href@noop {} {\bibfield  {journal} {\bibinfo  {journal} {J.
  Phys.-Paris}\ }\textbf {\bibinfo {volume} {43}},\ \bibinfo {pages} {197}
  (\bibinfo {year} {1982}{\natexlab{a}})}\BibitemShut {NoStop}%
\bibitem [{\citenamefont {Lhuillier}\ and\ \citenamefont
  {Lalo{\"e}}(1982{\natexlab{b}})}]{Lhuillier1982a}%
  \BibitemOpen
  \bibfield  {author} {\bibinfo {author} {\bibfnamefont {C.}~\bibnamefont
  {Lhuillier}}\ and\ \bibinfo {author} {\bibfnamefont {F.}~\bibnamefont
  {Lalo{\"e}}},\ }\href@noop {} {\bibfield  {journal} {\bibinfo  {journal} {J.
  Phys.-Paris}\ }\textbf {\bibinfo {volume} {43}},\ \bibinfo {pages} {225}
  (\bibinfo {year} {1982}{\natexlab{b}})}\BibitemShut {NoStop}%
\bibitem [{\citenamefont {Bashkin}(1981)}]{Bashkin1981}%
  \BibitemOpen
  \bibfield  {author} {\bibinfo {author} {\bibfnamefont {E.}~\bibnamefont
  {Bashkin}},\ }\href@noop {} {\bibfield  {journal} {\bibinfo  {journal} {JETP
  Lett+}\ }\textbf {\bibinfo {volume} {33}} (\bibinfo {year}
  {1981})}\BibitemShut {NoStop}%
\bibitem [{\citenamefont {Bashkin}(1984)}]{Bashkin1984}%
  \BibitemOpen
  \bibfield  {author} {\bibinfo {author} {\bibfnamefont {E.}~\bibnamefont
  {Bashkin}},\ }\href@noop {} {\bibfield  {journal} {\bibinfo  {journal} {Zh.
  Eksp. Teor. Fiz}\ }\textbf {\bibinfo {volume} {87}},\ \bibinfo {pages} {1948}
  (\bibinfo {year} {1984})}\BibitemShut {NoStop}%
\bibitem [{\citenamefont {Bashkin}(1986)}]{Bashkin1986}%
  \BibitemOpen
  \bibfield  {author} {\bibinfo {author} {\bibfnamefont {E.}~\bibnamefont
  {Bashkin}},\ }\href {http://stacks.iop.org/0038-5670/29/i=3/a=R02} {\bibfield
   {journal} {\bibinfo  {journal} {Sov. Phys. Usp.}\ }\textbf {\bibinfo
  {volume} {29}},\ \bibinfo {pages} {238} (\bibinfo {year} {1986})}\BibitemShut
  {NoStop}%
\bibitem [{\citenamefont {Lewandowski}\ \emph {et~al.}(2002)\citenamefont
  {Lewandowski}, \citenamefont {Harber}, \citenamefont {Whitaker},\ and\
  \citenamefont {Cornell}}]{Lewandowski2002}%
  \BibitemOpen
  \bibfield  {author} {\bibinfo {author} {\bibfnamefont {H.~J.}\ \bibnamefont
  {Lewandowski}}, \bibinfo {author} {\bibfnamefont {D.~M.}\ \bibnamefont
  {Harber}}, \bibinfo {author} {\bibfnamefont {D.~L.}\ \bibnamefont
  {Whitaker}}, \ and\ \bibinfo {author} {\bibfnamefont {E.~A.}\ \bibnamefont
  {Cornell}},\ }\href {\doibase 10.1103/PhysRevLett.88.070403} {\bibfield
  {journal} {\bibinfo  {journal} {Phys. Rev. Lett.}\ }\textbf {\bibinfo
  {volume} {88}},\ \bibinfo {pages} {070403} (\bibinfo {year}
  {2002})}\BibitemShut {NoStop}%
\bibitem [{\citenamefont {Du}\ \emph {et~al.}(2008)\citenamefont {Du},
  \citenamefont {Luo}, \citenamefont {Clancy},\ and\ \citenamefont
  {Thomas}}]{Du2008}%
  \BibitemOpen
  \bibfield  {author} {\bibinfo {author} {\bibfnamefont {X.}~\bibnamefont
  {Du}}, \bibinfo {author} {\bibfnamefont {L.}~\bibnamefont {Luo}}, \bibinfo
  {author} {\bibfnamefont {B.}~\bibnamefont {Clancy}}, \ and\ \bibinfo {author}
  {\bibfnamefont {J.~E.}\ \bibnamefont {Thomas}},\ }\href {\doibase
  10.1103/PhysRevLett.101.150401} {\bibfield  {journal} {\bibinfo  {journal}
  {Phys. Rev. Lett.}\ }\textbf {\bibinfo {volume} {101}},\ \bibinfo {pages}
  {150401} (\bibinfo {year} {2008})}\BibitemShut {NoStop}%
\bibitem [{\citenamefont {Du}\ \emph {et~al.}(2009)\citenamefont {Du},
  \citenamefont {Zhang}, \citenamefont {Petricka},\ and\ \citenamefont
  {Thomas}}]{Du2009b}%
  \BibitemOpen
  \bibfield  {author} {\bibinfo {author} {\bibfnamefont {X.}~\bibnamefont
  {Du}}, \bibinfo {author} {\bibfnamefont {Y.}~\bibnamefont {Zhang}}, \bibinfo
  {author} {\bibfnamefont {J.}~\bibnamefont {Petricka}}, \ and\ \bibinfo
  {author} {\bibfnamefont {J.~E.}\ \bibnamefont {Thomas}},\ }\href {\doibase
  10.1103/PhysRevLett.103.010401} {\bibfield  {journal} {\bibinfo  {journal}
  {Phys. Rev. Lett.}\ }\textbf {\bibinfo {volume} {103}},\ \bibinfo {pages}
  {010401} (\bibinfo {year} {2009})}\BibitemShut {NoStop}%
\bibitem [{\citenamefont {Bouchaud}\ and\ \citenamefont
  {Lhuillier}(1985)}]{Bouchaud1985}%
  \BibitemOpen
  \bibfield  {author} {\bibinfo {author} {\bibfnamefont {J.-P.}\ \bibnamefont
  {Bouchaud}}\ and\ \bibinfo {author} {\bibfnamefont {C.}~\bibnamefont
  {Lhuillier}},\ }\href@noop {} {\bibfield  {journal} {\bibinfo  {journal} {J.
  Phys.-Paris}\ }\textbf {\bibinfo {volume} {46}},\ \bibinfo {pages} {1101}
  (\bibinfo {year} {1985})}\BibitemShut {NoStop}%
\bibitem [{\citenamefont {Liu}\ \emph {et~al.}(2013)\citenamefont {Liu},
  \citenamefont {Pi{\'e}chon},\ and\ \citenamefont {Fuchs}}]{Liu2013b}%
  \BibitemOpen
  \bibfield  {author} {\bibinfo {author} {\bibfnamefont {Y.}~\bibnamefont
  {Liu}}, \bibinfo {author} {\bibfnamefont {F.}~\bibnamefont {Pi{\'e}chon}}, \
  and\ \bibinfo {author} {\bibfnamefont {J.}~\bibnamefont {Fuchs}},\
  }\href@noop {} {\bibfield  {journal} {\bibinfo  {journal} {Europhys. Lett.}\
  }\textbf {\bibinfo {volume} {103}},\ \bibinfo {pages} {17007} (\bibinfo
  {year} {2013})}\BibitemShut {NoStop}%
\bibitem [{\citenamefont {Kirsten}\ and\ \citenamefont
  {Toms}(1996)}]{Kirsten1996}%
  \BibitemOpen
  \bibfield  {author} {\bibinfo {author} {\bibfnamefont {K.}~\bibnamefont
  {Kirsten}}\ and\ \bibinfo {author} {\bibfnamefont {D.}~\bibnamefont {Toms}},\
  }\href@noop {} {\bibfield  {journal} {\bibinfo  {journal} {Phys. Lett. A}\
  }\textbf {\bibinfo {volume} {222}},\ \bibinfo {pages} {148} (\bibinfo {year}
  {1996})}\BibitemShut {NoStop}%
\bibitem [{\citenamefont {Miller~Jr}\ and\ \citenamefont
  {Good~Jr}(1953)}]{Miller1953}%
  \BibitemOpen
  \bibfield  {author} {\bibinfo {author} {\bibfnamefont {S.}~\bibnamefont
  {Miller~Jr}}\ and\ \bibinfo {author} {\bibfnamefont {R.}~\bibnamefont
  {Good~Jr}},\ }\href@noop {} {\bibfield  {journal} {\bibinfo  {journal} {Phys.
  Rev.}\ }\textbf {\bibinfo {volume} {91}},\ \bibinfo {pages} {174} (\bibinfo
  {year} {1953})}\BibitemShut {NoStop}%
\bibitem [{\citenamefont {Sakurai}\ \emph {et~al.}(1995)\citenamefont
  {Sakurai}, \citenamefont {Tuan},\ and\ \citenamefont
  {Commins}}]{Sakurai1995}%
  \BibitemOpen
  \bibfield  {author} {\bibinfo {author} {\bibfnamefont {J.}~\bibnamefont
  {Sakurai}}, \bibinfo {author} {\bibfnamefont {S.-F.}\ \bibnamefont {Tuan}}, \
  and\ \bibinfo {author} {\bibfnamefont {E.}~\bibnamefont {Commins}},\
  }\href@noop {} {\emph {\bibinfo {title} {Modern quantum mechanics, revised
  edition}}}\ (\bibinfo  {publisher} {AAPT},\ \bibinfo {year}
  {1995})\BibitemShut {NoStop}%
\bibitem [{\citenamefont {Schiff}(1968)}]{Schiff1968}%
  \BibitemOpen
  \bibfield  {author} {\bibinfo {author} {\bibfnamefont {L.}~\bibnamefont
  {Schiff}},\ }\href@noop {} {\emph {\bibinfo {title} {Quantum Mechanics
  3rd}}}\ (\bibinfo {year} {1968})\ pp.\ \bibinfo {pages} {61--62}\BibitemShut
  {NoStop}%
\bibitem [{\citenamefont {Walraven}(2010)}]{Walraven2010}%
  \BibitemOpen
  \bibfield  {author} {\bibinfo {author} {\bibfnamefont {J.}~\bibnamefont
  {Walraven}},\ }\href@noop {} {\bibfield  {journal} {\bibinfo  {journal} {Les
  Houches lectures, unpublished}\ } (\bibinfo {year} {2010})}\BibitemShut
  {NoStop}%
\bibitem [{\citenamefont {Van~Vleck}(1928)}]{VanVleck1928}%
  \BibitemOpen
  \bibfield  {author} {\bibinfo {author} {\bibfnamefont {J.}~\bibnamefont
  {Van~Vleck}},\ }\href@noop {} {\bibfield  {journal} {\bibinfo  {journal}
  {Proceedings of the National Academy of Sciences}\ }\textbf {\bibinfo
  {volume} {14}},\ \bibinfo {pages} {178} (\bibinfo {year} {1928})}\BibitemShut
  {NoStop}%
\bibitem [{\citenamefont {Schiller}(1962{\natexlab{a}})}]{Schiller1962}%
  \BibitemOpen
  \bibfield  {author} {\bibinfo {author} {\bibfnamefont {R.}~\bibnamefont
  {Schiller}},\ }\href@noop {} {\bibfield  {journal} {\bibinfo  {journal}
  {Phys. Rev.}\ }\textbf {\bibinfo {volume} {125}},\ \bibinfo {pages} {1109}
  (\bibinfo {year} {1962}{\natexlab{a}})}\BibitemShut {NoStop}%
\bibitem [{\citenamefont {Schiller}(1962{\natexlab{b}})}]{Schiller1962b}%
  \BibitemOpen
  \bibfield  {author} {\bibinfo {author} {\bibfnamefont {R.}~\bibnamefont
  {Schiller}},\ }\href@noop {} {\bibfield  {journal} {\bibinfo  {journal}
  {Phys. Rev.}\ }\textbf {\bibinfo {volume} {125}},\ \bibinfo {pages} {1100}
  (\bibinfo {year} {1962}{\natexlab{b}})}\BibitemShut {NoStop}%
\bibitem [{\citenamefont {Van~Horn}\ and\ \citenamefont
  {Salpeter}(1967)}]{VanHorn1967}%
  \BibitemOpen
  \bibfield  {author} {\bibinfo {author} {\bibfnamefont {H.}~\bibnamefont
  {Van~Horn}}\ and\ \bibinfo {author} {\bibfnamefont {E.}~\bibnamefont
  {Salpeter}},\ }\href@noop {} {\bibfield  {journal} {\bibinfo  {journal}
  {Phys. Rev.}\ }\textbf {\bibinfo {volume} {157}},\ \bibinfo {pages} {751}
  (\bibinfo {year} {1967})}\BibitemShut {NoStop}%
\bibitem [{\citenamefont {Sergeenko}(2000)}]{Sergeenko2000}%
  \BibitemOpen
  \bibfield  {author} {\bibinfo {author} {\bibfnamefont {M.}~\bibnamefont
  {Sergeenko}},\ }\href@noop {} {\bibfield  {journal} {\bibinfo  {journal}
  {Mod. Phys. Lett. A}\ }\textbf {\bibinfo {volume} {15}},\ \bibinfo {pages}
  {83} (\bibinfo {year} {2000})}\BibitemShut {NoStop}%
\bibitem [{\citenamefont {Gamble}\ and\ \citenamefont
  {Lindner}(2009)}]{Gamble2009}%
  \BibitemOpen
  \bibfield  {author} {\bibinfo {author} {\bibfnamefont {J.~K.}\ \bibnamefont
  {Gamble}}\ and\ \bibinfo {author} {\bibfnamefont {J.~F.}\ \bibnamefont
  {Lindner}},\ }\href@noop {} {\bibfield  {journal} {\bibinfo  {journal}
  {American Journal of Physics}\ }\textbf {\bibinfo {volume} {77}},\ \bibinfo
  {pages} {244} (\bibinfo {year} {2009})}\BibitemShut {NoStop}%
\bibitem [{\citenamefont {Bendat}\ and\ \citenamefont
  {Piersol}(2011)}]{Bendat2011}%
  \BibitemOpen
  \bibfield  {author} {\bibinfo {author} {\bibfnamefont {J.}~\bibnamefont
  {Bendat}}\ and\ \bibinfo {author} {\bibfnamefont {A.}~\bibnamefont
  {Piersol}},\ }\href@noop {} {\emph {\bibinfo {title} {Random data: analysis
  and measurement procedures}}},\ Vol.\ \bibinfo {volume} {729}\ (\bibinfo
  {publisher} {John Wiley \& Sons},\ \bibinfo {year} {2011})\BibitemShut
  {NoStop}%
\bibitem [{\citenamefont {Zweig}\ and\ \citenamefont
  {Hufnagel}(1990)}]{Zweig1990}%
  \BibitemOpen
  \bibfield  {author} {\bibinfo {author} {\bibfnamefont {D.}~\bibnamefont
  {Zweig}}\ and\ \bibinfo {author} {\bibfnamefont {R.}~\bibnamefont
  {Hufnagel}},\ }in\ \href@noop {} {\emph {\bibinfo {booktitle} {San Dieg-DL
  Tentative}}}\ (\bibinfo {organization} {International Society for Optics and
  Photonics},\ \bibinfo {year} {1990})\ pp.\ \bibinfo {pages}
  {295--302}\BibitemShut {NoStop}%
\bibitem [{\citenamefont {Larkin}(1996)}]{Larkin1996}%
  \BibitemOpen
  \bibfield  {author} {\bibinfo {author} {\bibfnamefont {K.~G.}\ \bibnamefont
  {Larkin}},\ }\href@noop {} {\bibfield  {journal} {\bibinfo  {journal} {J.
  Opt. Soc. Am. A}\ }\textbf {\bibinfo {volume} {13}},\ \bibinfo {pages} {832}
  (\bibinfo {year} {1996})}\BibitemShut {NoStop}%
\bibitem [{\citenamefont {Gibble}(2009)}]{Gibble2009}%
  \BibitemOpen
  \bibfield  {author} {\bibinfo {author} {\bibfnamefont {K.}~\bibnamefont
  {Gibble}},\ }\href {\doibase 10.1103/PhysRevLett.103.113202} {\bibfield
  {journal} {\bibinfo  {journal} {Phys. Rev. Lett.}\ }\textbf {\bibinfo
  {volume} {103}},\ \bibinfo {pages} {113202} (\bibinfo {year}
  {2009})}\BibitemShut {NoStop}%
\bibitem [{\citenamefont {Maineult}\ \emph {et~al.}(2012)\citenamefont
  {Maineult}, \citenamefont {Deutsch}, \citenamefont {Gibble}, \citenamefont
  {Reichel},\ and\ \citenamefont {Rosenbusch}}]{Maineult2012}%
  \BibitemOpen
  \bibfield  {author} {\bibinfo {author} {\bibfnamefont {W.}~\bibnamefont
  {Maineult}}, \bibinfo {author} {\bibfnamefont {C.}~\bibnamefont {Deutsch}},
  \bibinfo {author} {\bibfnamefont {K.}~\bibnamefont {Gibble}}, \bibinfo
  {author} {\bibfnamefont {J.}~\bibnamefont {Reichel}}, \ and\ \bibinfo
  {author} {\bibfnamefont {P.}~\bibnamefont {Rosenbusch}},\ }\href {\doibase
  10.1103/PhysRevLett.109.020407} {\bibfield  {journal} {\bibinfo  {journal}
  {Phys. Rev. Lett.}\ }\textbf {\bibinfo {volume} {109}},\ \bibinfo {pages}
  {020407} (\bibinfo {year} {2012})}\BibitemShut {NoStop}%
\bibitem [{\citenamefont {Bigelow}\ \emph {et~al.}(1989)\citenamefont
  {Bigelow}, \citenamefont {Freed},\ and\ \citenamefont {Lee}}]{Bigelow1989}%
  \BibitemOpen
  \bibfield  {author} {\bibinfo {author} {\bibfnamefont {N.~P.}\ \bibnamefont
  {Bigelow}}, \bibinfo {author} {\bibfnamefont {J.~H.}\ \bibnamefont {Freed}},
  \ and\ \bibinfo {author} {\bibfnamefont {D.~M.}\ \bibnamefont {Lee}},\
  }\href@noop {} {\bibfield  {journal} {\bibinfo  {journal} {Phys. Rev. Lett.}\
  }\textbf {\bibinfo {volume} {63}},\ \bibinfo {pages} {1609} (\bibinfo {year}
  {1989})}\BibitemShut {NoStop}%
\bibitem [{\citenamefont {Reif}(1983)}]{Reif1965}%
  \BibitemOpen
  \bibfield  {author} {\bibinfo {author} {\bibfnamefont {F.}~\bibnamefont
  {Reif}},\ }\href@noop {} {\emph {\bibinfo {title} {Fundamentals of
  statistical and thermal physics}}}\ (\bibinfo  {publisher} {McGraw-Hill},\
  \bibinfo {year} {1983})\BibitemShut {NoStop}%
\bibitem [{\citenamefont {Bradley}\ and\ \citenamefont
  {Gardiner}(2002)}]{Bradley2002}%
  \BibitemOpen
  \bibfield  {author} {\bibinfo {author} {\bibfnamefont {A.~S.}\ \bibnamefont
  {Bradley}}\ and\ \bibinfo {author} {\bibfnamefont {C.~W.}\ \bibnamefont
  {Gardiner}},\ }\href {http://stacks.iop.org/0953-4075/35/i=20/a=315}
  {\bibfield  {journal} {\bibinfo  {journal} {J. Phys. B-At. Mol. opt.}\
  }\textbf {\bibinfo {volume} {35}},\ \bibinfo {pages} {4299} (\bibinfo {year}
  {2002})}\BibitemShut {NoStop}%
\bibitem [{\citenamefont {Gardiner}\ and\ \citenamefont
  {Zoller}(1997)}]{Gardiner1997}%
  \BibitemOpen
  \bibfield  {author} {\bibinfo {author} {\bibfnamefont {C.}~\bibnamefont
  {Gardiner}}\ and\ \bibinfo {author} {\bibfnamefont {P.}~\bibnamefont
  {Zoller}},\ }\href@noop {} {\bibfield  {journal} {\bibinfo  {journal} {Phys.
  Rev. A}\ }\textbf {\bibinfo {volume} {55}},\ \bibinfo {pages} {2902}
  (\bibinfo {year} {1997})}\BibitemShut {NoStop}%
\bibitem [{\citenamefont {Fuchs}\ \emph {et~al.}(2002)\citenamefont {Fuchs},
  \citenamefont {Gangardt},\ and\ \citenamefont {Lalo{\"e}}}]{Fuchs2002}%
  \BibitemOpen
  \bibfield  {author} {\bibinfo {author} {\bibfnamefont {J.-N.}\ \bibnamefont
  {Fuchs}}, \bibinfo {author} {\bibfnamefont {D.}~\bibnamefont {Gangardt}}, \
  and\ \bibinfo {author} {\bibfnamefont {F.}~\bibnamefont {Lalo{\"e}}},\
  }\href@noop {} {\bibfield  {journal} {\bibinfo  {journal} {Phys. Rev. Lett.}\
  }\textbf {\bibinfo {volume} {88}},\ \bibinfo {pages} {230404} (\bibinfo
  {year} {2002})}\BibitemShut {NoStop}%
\bibitem [{\citenamefont {Solaro}(2016)}]{Solaro2016b}%
  \BibitemOpen
  \bibfield  {author} {\bibinfo {author} {\bibfnamefont {C.}~\bibnamefont
  {Solaro}},\ }\emph {\bibinfo {title} {Interf{\'e}rom{\`e}tres atomiques
  pi{\'e}g{\'e}s: du r{\'e}gime dilu{\'e} au r{\'e}gime dense}},\ \href@noop {}
  {Ph.D. thesis},\ \bibinfo  {school} {Universit{\'e} Pierre et Marie
  Curie-Paris VI} (\bibinfo {year} {2016})\BibitemShut {NoStop}%
\end{thebibliography}%

\end{document}